\documentclass[aps,prx,longbibliography,superscriptaddress,twocolumn,showpacs]{revtex4-2}

\usepackage[utf8]{inputenc}
\usepackage[american,british]{babel}
\usepackage[T1]{fontenc}

\usepackage{amssymb}

\usepackage{graphicx}
\usepackage{xcolor}
\usepackage{dcolumn}
\usepackage{bm}
\usepackage[normalem]{ulem}

\usepackage{amsmath,amsthm,amssymb,mathrsfs,mathtools,amsfonts,braket}
\usepackage[caption=false]{subfig}

\usepackage{braket}
\usepackage{todonotes}
\newcounter{todocounter}
\usepackage[Option]{overpic} 
\usepackage{mathdots}
\usepackage{dsfont}
\usepackage{mathrsfs}
\DeclareMathAlphabet{\mathscrbf}{OMS}{mdugm}{b}{n}

 \newcommand{\RN}[1]{
  \textup{\expandafter{(\romannumeral#1)}}%
}
\usepackage{comment}

\usepackage{bigints}
\usepackage{float}
\usepackage{scalerel,stackengine,amsmath}

\begin{document}

~
\title{An efficient method for quantum impurity problems out of equilibrium}

\author{Julian Thoenniss}
\thanks{These authors contributed equally to this work.}
\affiliation{Department of Theoretical Physics,
University of Geneva, Quai Ernest-Ansermet 30,
1205 Geneva, Switzerland}

\author{Michael  Sonner}
\thanks{These authors contributed equally to this work.}
\affiliation{Department of Theoretical Physics,
University of Geneva, Quai Ernest-Ansermet 30,
1205 Geneva, Switzerland}

\author{Alessio Lerose}
\affiliation{Department of Theoretical Physics,
University of Geneva, Quai Ernest-Ansermet 30,
1205 Geneva, Switzerland}

\author{Dmitry A. Abanin}
\affiliation{Department of Theoretical Physics,
University of Geneva, Quai Ernest-Ansermet 30,
1205 Geneva, Switzerland}

\date{\today}

\begin{abstract}

We introduce an efficient method to simulate dynamics of an interacting quantum impurity coupled to non-interacting fermionic reservoirs. Viewing the impurity as an open quantum system, we describe the reservoirs by their Feynman-Vernon influence functionals (IF). The IF are represented as matrix-product states in the temporal domain, which enables an efficient computation of  dynamics for arbitrary  interactions. We apply our method to study quantum quenches and transport in an Anderson impurity model, including highly non-equilibrium setups, and find favorable performance compared to state-of-the-art methods. The computational resources required for an accurate computation of dynamics scale polynomially with evolution time, indicating that a broad class of out-of-equilibrium quantum impurity problems are efficiently solvable. This approach will provide new insights into dynamical properties of mesoscopic devices and correlated materials. 

\end{abstract}

\maketitle



{\bf Introduction.} Non-equilibrium many-body dynamics is actively investigated in condensed matter and synthetic quantum systems such as ultracold atoms~\cite{BlochColdAtoms}. The aim of the ongoing quest is to find regimes where a non-equilibrium system exhibits desired physical properties, which may be qualitatively different compared to equilibrium. Theoretically, out-of-equilibrium many-body problems are extremely challenging, both for analytical and numerical methods~\cite{PAECKEL2019167998,aoki14neqdmftreview}. 

Quantum impurity models (QIM), where a small quantum system such as a quantum dot is coupled to reservoir(s) of itinerant electrons, naturally arise in a variety of systems, including mesoscopic conductors~\cite{GlazmanKondo2004} and ultracold atoms~\cite{KanaszNagy18ExploringKondo,Riegger18LocalizedMagneticMoments}. Even relatively simple QIM such as the celebrated Anderson impurity model (AIM)~\cite{AIM}, exhibit rich many-body physics including the Kondo effect whereby the impurity spin is screened by itinerant electrons~\cite{hewson_1993}. Fermionic QIM, including the Anderson models, also play a central role in state-of-the-art methods for strongly correlated materials such as dynamical mean-field theory (DMFT), where the material properties are expressed via a self-consistent QIM~\cite{GeorgesRMP,aoki14neqdmftreview}. 

A large number of methods for non-equilibrium QIM, and in particular for the AIM, have been developed in recent years. These include iterative path-integral approximations~\cite{MakriMakarov94,EggerIterative2008,MillisImp2010}, non-Markovian~\cite{Tu08nonmarkovian,IFnanodevices} or auxiliary master equations (AME)~\cite{dorda14auxiliary,Lotem20renormalized}, hierarchical equations of motion (HEOM)~\cite{Tanimura89,Jin08heom,Dan22efficientheom}, time-dependent numerical renormalization group (NRG)~\cite{Anders05Realtime,nghiem17timeevolution,DelftNonEquil18} and density matrix renormalization group (tDMRG)~\cite{prior10efficient,Nuss15SpatiotemporalFormation,WolfPRB14,Nusseler20Efficient,Wojtowicz20OpensystemTN,Kohn_2022}, various variants of Quantum Monte Carlo (QMC)~\cite{Muhlbacher08Realtime,Schiro09realtime,Werner09Diagrammatic,gull11NumericallyExact,cohen11memory,cohen13neqkondo}, as well as variational~\cite{Ashida2018,SHI2018245} techniques. 
Recent advances including inchworm algorithm~\cite{cohen15taming} and increasingly sophisticated high-order diagrammatic calculations~\cite{Bertrand19Reconstructing,QQMC20} ameliorated the sign problem of QMC, thereby giving access to longer evolution times.
However, despite recent developments, the current methods cannot provide guarantees of computational efficiency for out-of-equilibrium QIM, which remain a subject of active research. 

\begin{figure}[t]
\includegraphics[width=0.48\textwidth]{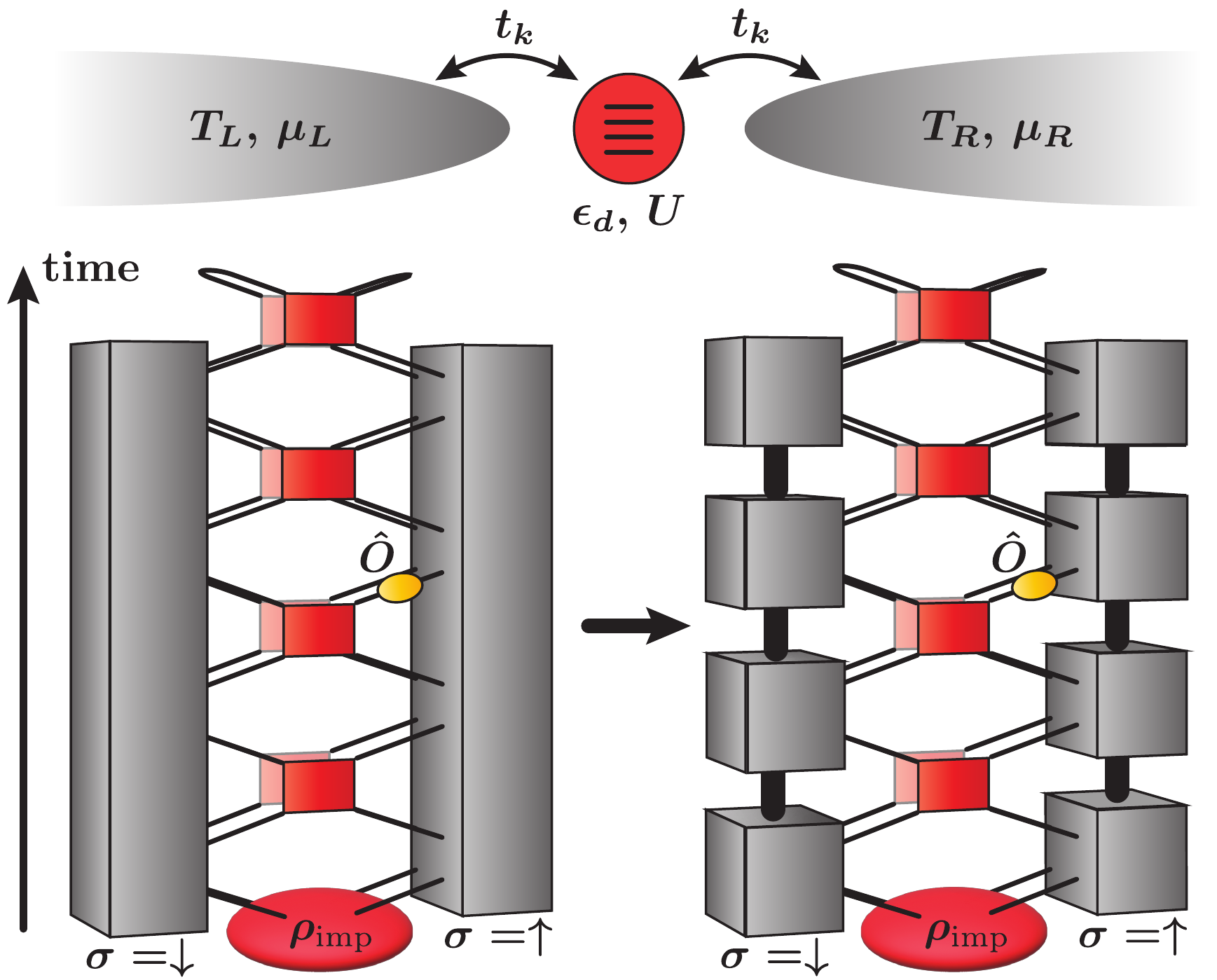}
\centering
\caption{
Top: Illustration of single impurity Anderson model [Eq.~\eqref{eq:SIAM_Hamiltonian}] with an impurity (red) tunnel-coupled to two reservoirs (gray). 
Bottom: Tensor-network representation of a time-dependent observable $\langle \hat{O}(t)\rangle$. 
The dynamical influence of the environment is encoded in a single IF per orbital degree of freedom (here, two gray tensors for $\sigma = \uparrow, \downarrow$,  left) which can be efficiently represented as MPS in the temporal domain (right) and hence contracted with the local impurity evolution (product of red tensors). Foreground [background] layer represents forward [backward] branch of the  Keldysh contour.
}
\label{fig:setup}
\end{figure}

In this Article, we present a conceptually simple and efficient method for fermionic QIM, building on recent developments in describing interacting~\cite{Banuls09,huang14longtime,lerose2020,lerose2021,Chan21,sonner2021influence,sonner22characterizing,Piroli2020,klobas2021exact,giudice2021temporal,lerose2022overcoming} and non-interacting~\cite{TEMPO,Thoenniss22,bose2021tensor} quantum baths using temporal tensor networks.
The starting point of our approach is to treat the impurity as an open quantum system coupled to the ``bath'' that consists of fermionic leads (Fig.~1). The effect of the leads is then represented by the fermionic extension of the Feynman-Vernon influence functional (IF)~\cite{FeynmanVernon}, which can be obtained in closed form for arbitrary non-interacting reservoirs~\cite{aoki14neqdmftreview,IFnanodevices,Thoenniss22}.
As a key ingredient of our approach, the IF can be efficiently represented as a matrix-product state (MPS) in the temporal domain with controlled bond dimension, thanks to the favorable scaling of temporal entanglement of the IF~\cite{lerose2021,Thoenniss22}. This enables an efficient computation of time-dependent observables at the impurity location (e.g. charge, spin, currents) via straightforward tensor contraction.

We demonstrate the efficiency of our method for paradigmatic non-equilibrium QIM setups, including (i) a quantum quench, where impurity site is connected to equilibrium leads at time $t=0$ and (ii) a biased AIM with two imbalanced leads. In all cases, our method is capable of reproducing and going beyond the state-of-the-art results obtained by inchworm and diagrammatic QMC.


Besides conceptual simplicity, the method presented here has a number of advantages. First and foremost, required resources grow polynomially   in evolution time. In terms of computational complexity~\cite{bravyi2017complexity,Debertolis21fewbody}, this implies that QIM are efficiently solvable even far away from equilibrium.
Furthermore, the method is non-perturbative, in contrast e.g. to QMC, which involves perturbative expansions  either in the impurity-reservoirs hybridization 
or in the on-site Coulomb interaction. In addition, from a practical viewpoint, once an efficient MPS representation of the reservoirs' IF is found, dynamics of impurities with an arbitrary choice of time-dependent local Hamiltonian can be subsequently computed with modest effort. 


{\bf Description of the method.} 
We consider the single-impurity Anderson model, described by the Hamiltonian
\begin{equation}
H =  \sum_{\substack {k\\\sigma=\uparrow,\downarrow\\\alpha=L,R}} \Big[ \big( t_k d_\sigma^\dagger c_{k,\alpha,\sigma} + h.c.\big) + \epsilon_{k} c_{k,\alpha, \sigma}^\dagger c_{k,\alpha,\sigma}\Big] + H_\text{imp},
    \label{eq:SIAM_Hamiltonian}
\end{equation} with $H_\text{imp} =  (\epsilon_{d} - U/2) \sum_\sigma  \hat d_\sigma^\dagger \hat d_\sigma 
+ U  \hat d^\dagger_{\uparrow}\hat d_{\uparrow} \hat d^\dagger_{\downarrow} \hat d_{\downarrow} .$
The impurity level described by fermions $d_\sigma$ is coupled to two baths ($\alpha=L,R$) of free fermions $c_{k,\alpha,\sigma}$ with identical  dispersion $\epsilon_k$ and tunnel couplings $t_k$, initially in thermal equilibrium (see top illustration in Fig.~\ref{fig:setup}).
 Coulomb interaction $U\neq 0$ in $H_\text{imp}$  gives rise to strong correlations in and out of equilibrium.

We are primarily interested in the real-time evolution of an impurity observable $\langle \hat{O}(t)\rangle$ starting from a factorized initial state $\rho(0)=\rho_L \otimes \rho_{\text{imp}} \otimes \rho_R$, with $\rho_{L,R}$ equilibrium states at inverse temperatures $\beta_{L,R}$ and chemical potentials $\mu_{L,R}$.
While conventional tensor-network approaches attempt to compactly represent $\rho(t)$~\cite{PAECKEL2019167998}, we instead express $\langle \hat{O}(t)\rangle$ as a Keldysh path integral over Grassmann trajectories of impurity and baths.  Gaussian integration over the bath trajectories gives
\begin{multline}
\langle \hat{O}(t)\rangle
\;\propto\;
\int \bigg(\prod_{\sigma,\tau} d\bar{\eta}_{\sigma,\tau}d\eta_{\sigma,\tau}\bigg) \mathcal{O}(\boldsymbol{\bar{\eta}}_{t},\boldsymbol{\eta}_{t}) \\
\times \exp\bigg\{\int_\mathcal{C} d\tau  \Big[\sum_\sigma \bar{\eta}_{\sigma,\tau}\partial_{\tau} \eta_{\sigma,\tau}
-i \mathcal{H}_\text{imp}(\boldsymbol{\bar{\eta}}_{\tau},\boldsymbol{\eta}_{\tau})\Big] \bigg\}\\
\times \mathcal{\rho}_\text{imp}[\bar{\bm{\eta}}_{0},\bm{\eta}_{0}]\,\prod_{\sigma=\uparrow,\downarrow} \exp\bigg(\int_\mathcal{C}d\tau \int_\mathcal{C} d\tau^\prime \bar{\eta}_{\sigma,\tau} \Delta(\tau,\tau^\prime) \eta_{\sigma,\tau^\prime} \bigg).
\label{eq:expec_value}
\end{multline}
Here $\boldsymbol{\bar{\eta}}_{\tau}=({\bar{\eta}}_{\uparrow,\tau},{\bar{\eta}}_{\downarrow,\tau})$ and $\boldsymbol{{\eta}}_{\tau}=({{\eta}}_{\uparrow,\tau},{{\eta}}_{\downarrow,\tau})$ parametrize the impurity trajectory. The IF is the last exponential in Eq.~(\ref{eq:expec_value}), defined by the hybridization function $\Delta(\tau,\tau^\prime)=\sum_\alpha \Delta^\alpha(\tau,\tau^\prime)$, where $\Delta^\alpha$ fully encodes 
the dynamical influence of the bath $\alpha$,
\begin{equation}
\Delta^\alpha(\tau,\tau^\prime) =  \int \frac{d\omega}{2\pi} \Gamma(\omega) g^\alpha_{\tau,\tau^\prime}(\omega).
\label{eq:hyb_func}
\end{equation}
The latter is determined by the bath's spectral density
$\Gamma(\omega) =  2\pi \sum_k |t_{k}|^2 \delta(\omega - \epsilon_k)$ {and non-interacting Green’s function} $g^\alpha_{\tau,\tau^\prime}(\omega) = \big( n^\alpha_\text{F}(\omega) -\Theta_\mathcal{C}(\tau,\tau^\prime)\big) e^{-i\omega (\tau-\tau^\prime)}$, where $n_\text{F}^\alpha$ is the Fermi distribution at inverse temperature $\beta_\alpha$ and chemical potential~$\mu_\alpha$ and $\Theta_\mathcal{C}$ is the Heaviside step function on the Keldysh contour $\mathcal{C}$ (see e.g. Ref.~\cite{aoki14neqdmftreview}). Equation~(\ref{eq:expec_value}) is the starting point of advanced techniques for impurity dynamics such as AME, HEOM or QMC.


The difficulty in evaluating the path integral arises from the combination of non-Gaussianity (in $\mathcal{H}_\text{imp}$) and time-non-locality (in $\Delta(\tau,\tau^\prime)$). 
The key idea of our method is to interpret Eq.~(\ref{eq:expec_value}) as a scalar product of fictitious states and operators defined in a fermionic Fock space on a temporal lattice.
To that end,  we note that the textbook expression in Eq.~(\ref{eq:expec_value}) is defined as the limit $M\to\infty$ of a discrete-time expression, obtained by dividing the full time evolution window $[0,T]$ into $M$ steps of size $\delta t = T/M$; 
we fix a sufficiently large $M$. 
For our purpose, it is convenient to use a Trotter scheme that further splits the Trotter step into impurity and hybrization, leading to $8M$ trajectory variables per spin species along the discretized Keldysh contour, see Supplemental Material (SM) for details. We arrange these in two arrays, $\bm{\eta}_\sigma= (\eta_{\sigma,0^+},\eta_{\sigma,0^-},\hdots,\eta_{\sigma,(2M-1)^+},\eta_{\sigma,(2M-1)^-})$ and analogously $\bar{\bm{\eta}}_\sigma$, with degrees of freedom alternating on the forward ($+$) and backward ($-$) branch of the Keldysh contour. 
A series of manipulations with the discrete-time path integral, including partial ``particle-hole transformations'' $\eta \leftrightarrow \bar\eta$, allows us to rewrite Eq.~(\ref{eq:expec_value}) in a scalar product form (see SM for details):
\begin{align}\nonumber
\langle \hat{O}(t)\rangle \; \propto \; & \int \bigg(\prod_{\sigma} d\bar{\bm{\eta}}_{\sigma}d\bm{\eta}_{\sigma}\bigg) \\ \nonumber
&\times
\mathcal{I}[\bm{\eta}_\downarrow]e^{-\bar{\bm{\eta}}_\downarrow \bm{\eta}_\downarrow} \mathcal{D}_{\mathcal{O},t}[\bar{\bm{\eta}}_\downarrow,\bm{\eta}_\uparrow]\, 
e^{-\bar{\bm{\eta}}_\uparrow \bm{\eta}_\uparrow}
\mathcal{I}[\bar{\bm{\eta}}_{\uparrow}]\\
\equiv \; & \bra{I }\hat{D}_{\hat{O},t} \ket{I}.
\label{eq:expec_value_overlap}
\end{align}
Here, the kernel
$\mathcal{D}_{\mathcal{O},t}[\bar{\bm{\eta}}_\downarrow,\bm{\eta}_\uparrow]$,
which is non-Gaussian, describes impurity's own dynamics, and has a simple
product form due to time locality. This gives rise to a product operator
{$\hat{D}_{{\hat{O},t}}=\hat D_1\otimes \cdots \otimes \hat D_M$}, where each
$\hat D_{m}$ is a $16\times16$ matrix (except the first and last: see superimposed red tensors in
Fig.~\ref{fig:setup}) and $\hat D_{m^*=t/\delta t}$ contains $\hat O$. The
discrete-time IF has a Gaussian form, $\mathcal{I}[\bm{\eta}_\sigma] =
\exp\big( \bm{\eta}_\sigma^T \,\mathbf{B}\,\bm{\eta}_\sigma\big)$,
where the antisymmetric matrix $\mathbf{B}$ is related to the time-discretization of $\Delta(\tau,\tau^\prime)$ (see SM).
The Gaussian many-body wave function $\ket{I}$ associated with $\mathcal{I}$ (gray tensors in Fig.~\ref{fig:setup} bottom left) is obtained by replacing Grassmann variables by corresponding creation operators acting on the Fock space vacuum, $\bm{c}^\dagger\equiv (c^\dagger_{0^+},c^\dagger_{0^-},\hdots,c^\dagger_{(2M-1)^+},c^\dagger_{(2M-1)^-})$,
\begin{equation}
    \ket{I} = \exp\Big( {\mathbf{c}^\dagger}^T\mathbf{B}\, \mathbf{c}^\dagger\Big)\ket{\emptyset}.
    \label{eq_Istate}
\end{equation}
Such a state formally has a Bardeen-Cooper-Schrieffer form, regardless of the fermion-number conservation of the original problem, cf. Eq.~\eqref{eq:expec_value}; this  is related to the ``particle-hole transformations'' performed to arrive at Eq.~(\ref{eq:expec_value_overlap}). We note that particle number conservation shows up as a sublattice symmetry in Eq.~\eqref{eq_Istate}. 

Next, we aim to represent the state $\ket{I}$ as a MPS. Correlations of this state, described by the function $\Delta(\tau,\tau^\prime)$, reflect non-Markovianity of the bath. 
The possibility of a compact MPS representation is determined by the entanglement properties of a wave function; we previously showed~\cite{Thoenniss22} that Gaussian IF wave functions arising in QIM exhibit at most logarithmic scaling of temporal entanglement with evolution time for both equilibrium and certain non-equilibrium initial states of the reservoirs. This suggests that such wave functions can be described by a polynomial-in-$T$ number of parameters.

Previous works~\cite{Fishman2015MPS,SchuchGaussianMPS} proposed algorithms for representing a fermionic Gaussian wave function as a MPS. Here we apply the Fishman-White (FW)  algorithm~\cite{Fishman2015MPS}, extended to BCS-like wave functions~\cite{Thoenniss22}. We first approximately represent the Gaussian state determined by~$\mathbf{B}$ [Eq.~(\ref{eq_Istate})] as a quantum circuit of nearest-neighbor Gaussian unitary gates applied to the vacuum (a product state in a temporal chain of {$4M$} spins). 
The approximation is controlled by a threshold parameter $\epsilon$ of the algorithm~\cite{Fishman2015MPS,Thoenniss22}, which determines the maximum number $D$ of gates acting on a given site in this circuit (which we refer to as ``local depth'' below).  Second, we compress the circuit with standard singular-value truncations to produce a MPS approximation of $\ket{I}$ with bond dimension $\chi \le 2^D$. 
Once the MPS is obtained (gray tensors in Fig.~\ref{fig:setup} bottom right), the impurity's reduced density matrix time evolved with an arbitrary (possibly time-dependent) impurity Hamiltonian $H_{\text{imp}}$ can be efficiently computed by tensor contraction in the time direction.
This method is straightforwardly applicable to the computation of multi-time observables, e.g. the impurity Green's function, as well as currents (see below). 

\begin{figure}[t]
\includegraphics[width=0.48\textwidth]{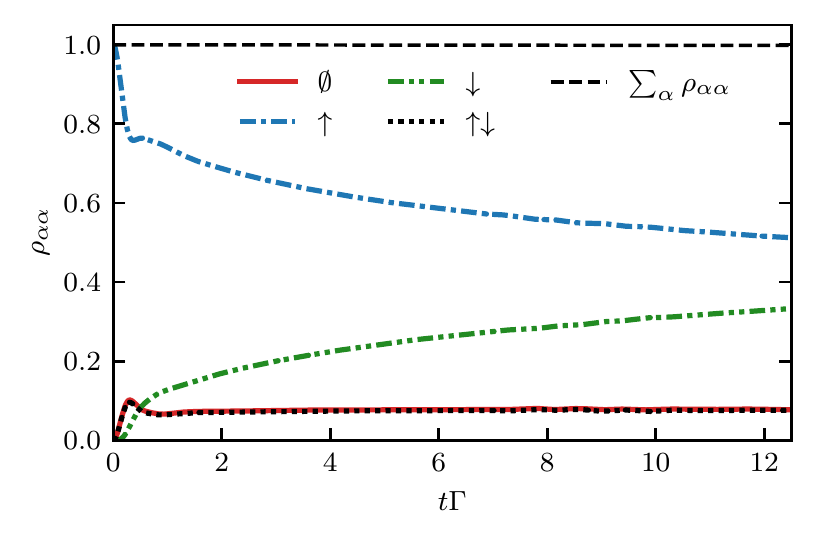}
\centering
\caption{Real-time evolution of the impurity density matrix after a quench. The plot reports diagonal entries $\rho_{\alpha\alpha},$ with $\alpha = \emptyset,\uparrow,\downarrow,\uparrow\downarrow$ as a function of time. The environment is modelled as in Ref.~\cite{cohen15taming} (see main text), with $\beta=50/\Gamma$ and $\mu=0$. Simulation parameters: Bond dimension $\chi=256$ per spin species, FW threshold  $\epsilon=5\cdot 10^{-13}$, Trotter step $ \delta t  = 0.02/\Gamma$.
}
\label{fig:Cohen_benchmark}
\end{figure}

{\bf A quantum quench.} 
As a first application of our method, we study a local quantum quench, where tunneling between impurity and the bath -- initially in equilibrium at equal $\beta$ and $\mu$ -- is turned on at time $t=0$. We monitor the real-time evolution of the impurity level population at $t>0$. 
In the Kondo regime  (strong interaction and low temperature), strong correlations develop in real time between the impurity and the bath, corresponding to the formation of a local screening cloud over a non-perturbatively long timescale -- a real-time manifestation of the Kondo effect, which was previously investigated with other methods~\cite{Anders05Realtime,Medvedyeva13SpatiotemporalBuildup,Nuss15SpatiotemporalFormation,nghiem17timeevolution,Ashida2018}.

Here we benchmark the state-of-the-art results of inchworm QMC in Ref.~\cite{cohen15taming}: We consider a bath defined by a flat band with smooth edges, $\Gamma(\omega)= \Gamma/\big[(1+e^{\nu(\omega - \omega_c)})(1+e^{-\nu(\omega + \omega_c)})\big]$ with $\omega_c = 10\Gamma$ and $\nu =10/\Gamma$. Moreover, we set $\beta=50/\Gamma$, $\mu=0$. We prepare the impurity in a singly occupied state $\rho_{\text{imp}}= \ket{\uparrow}\bra{\uparrow}$, with $\epsilon_d=0$ and $U=8\Gamma$, and couple it to the bath at time $t=0$. In Fig.~\ref{fig:Cohen_benchmark} we report our results for the evolution of the diagonal components of the impurity's reduced density matrix. Data are converged with respect to all simulation parameters (see caption), demonstrating accuracy beyond the data of Ref.~\cite{cohen15taming}. These results showcase the ability of our method to capture the slow dynamical formation of a spin singlet in the Kondo regime, which will be further investigated elsewhere.

{\bf Non-equilibrium transport.}
The system described by Eq.~\eqref{eq:SIAM_Hamiltonian} with a temperature or
chemical potential bias between $L$ and $R$ reservoirs models paradigmatic
non-equilibrium setups with correlated nanodevices. Capturing the full
transient charge and spin dynamics after a quench (either of tunnel-couplings
or of interactions) toward the non-equilibrium stationary state is a recurrent
challenging test for novel advanced numerical
techniques~\cite{schmidt08transient,werner10weakcoupling,Fugger_2018,DelftNonEquil18,Bertrand19Reconstructing}.

\begin{figure}[t!]
\includegraphics[width=0.48\textwidth]{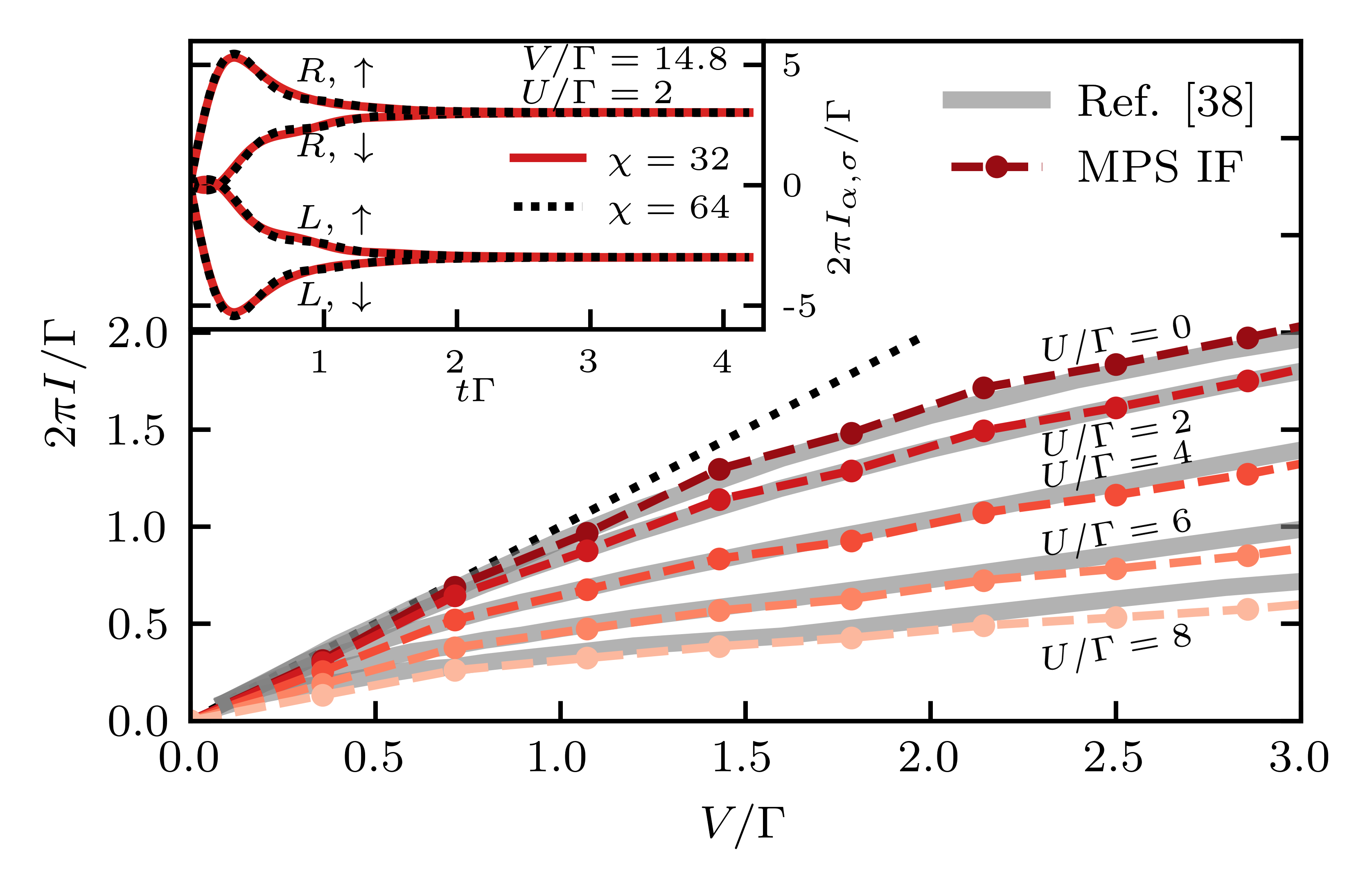}
\centering
\caption{ Current-voltage characteristics of an AIM.  Reservoirs $L$ and $R$ are tight-binding chains as in Refs.~\cite{Bertrand19Reconstructing,Nuss15SpatiotemporalFormation} (see main text), of $L=600$ sites each, at zero temperature and chemical potentials $\pm V/2$. Simulation parameters: Bond dimension $\chi=32$ per reservoir per spin species, FW threshold  $\epsilon=1\cdot 10^{-12}$, Trotter step $ \delta t = 0.007/\Gamma$.
For all values of $V$ and $U$ we evolve until time $T =4.2/\Gamma$ {and verify that at this time stationary state is reached}. Inset: At fixed $V/\Gamma = 14.8$ and $U/\Gamma = 2,$ we demonstrate convergence in bond dimension for all four components of the transient current, $\langle I_{\alpha,\sigma}(t) \rangle$ with $\alpha=L,R$ and $\sigma = \uparrow,\downarrow.$ 
}
\label{fig:current}
\end{figure}

Here we benchmark the state-of-the-art computation of the system's current-voltage
characteristics in
Ref.~\cite{Bertrand19Reconstructing}. We model the reservoirs as two homogenous tight-binding chains with
 nearest-neighbor hopping $t_{\rm hop}=1$, coupled to the impurity with tunneling amplitude
$t'_{\rm hop}=0.3162$, corresponding to a resonance width $\Gamma(\epsilon_d=0)=0.1$ (cf. Ref.~\cite{Nuss15SpatiotemporalFormation}). We initialize the two
reservoirs at zero temperature and chemical potentials $\pm V/2$, and monitor the time-dependent  current flowing through the impurity for several values of $U$, until the stationary state is reached. 

Unlike the contraction illustrated in Fig.~\ref{fig:setup} and used above for
the quench simulation, computing the current into either reservoir requires one to
keep track of the separate influence of reservoirs $L$ and $R$. A suitable
Trotter decomposition (see Ref.~\cite{Thoenniss22} and SM) allows us to couple
the two reservoirs with the impurity alternatively in discrete time steps
$\delta t$. The current of spin $\sigma$ electrons flowing into reservoir
$\alpha$, can then be computed as $\langle I_{\alpha,\sigma}(t)\rangle=\frac 1
{\delta t} \big[ \langle  d^\dagger_\sigma (t+\delta t) d_\sigma(t+\delta t)
\rangle - \langle d^\dagger_\sigma (t) d_\sigma(t) \rangle \big]$, where the
impurity interacts only with reservoir $\alpha$ during the time step from $t$
to $t+\delta t$. 

Keeping track of $L$ and $R$ separately results in a tensor contraction with
four IF MPS. This considerably limits the bond dimension we can afford for each
IF, as the final impurity evolution entails storing matrices acting on a $16
\chi^4$-dimensional space (while it was $16 \chi^2$ before). Nonetheless, we
found that the value of the current is converged over the full transient to the
stationary state for bond dimension as low as $\chi=32$ (see inset of
Fig.~\ref{fig:current}).

Figure~\ref{fig:current} shows the results of our computations, as well as the
corresponding data from Fig.~15 of Ref.~\cite{Bertrand19Reconstructing}. We
find a fairly good agreement throughout the wide explored parameter regime. The
unit slope of the dotted line represents the universal Landauer linear-response
conductance, $I=(e^2/h)V$ (recall $e=\hbar=1$ in our units). We note that small
discrepancies are to be expected at large biases $ V \gg \Gamma$ due to
non-universal effects of finite bath bandwidth ($t_{\rm hop}=10 \Gamma$ here).
We further remark that for small bias and large interaction the non-equilibrium
Kondo regime is approached, characterized by slow relaxation. Accordingly, in
the computation with smallest bias $V=0.36\Gamma$ and largest interaction
$U=8\Gamma$ in Fig.~\ref{fig:current}, the time-dependent current has not yet
fully reached its stationary value at time $T$.

\begin{figure}[t!]
\includegraphics[width=0.46\textwidth]{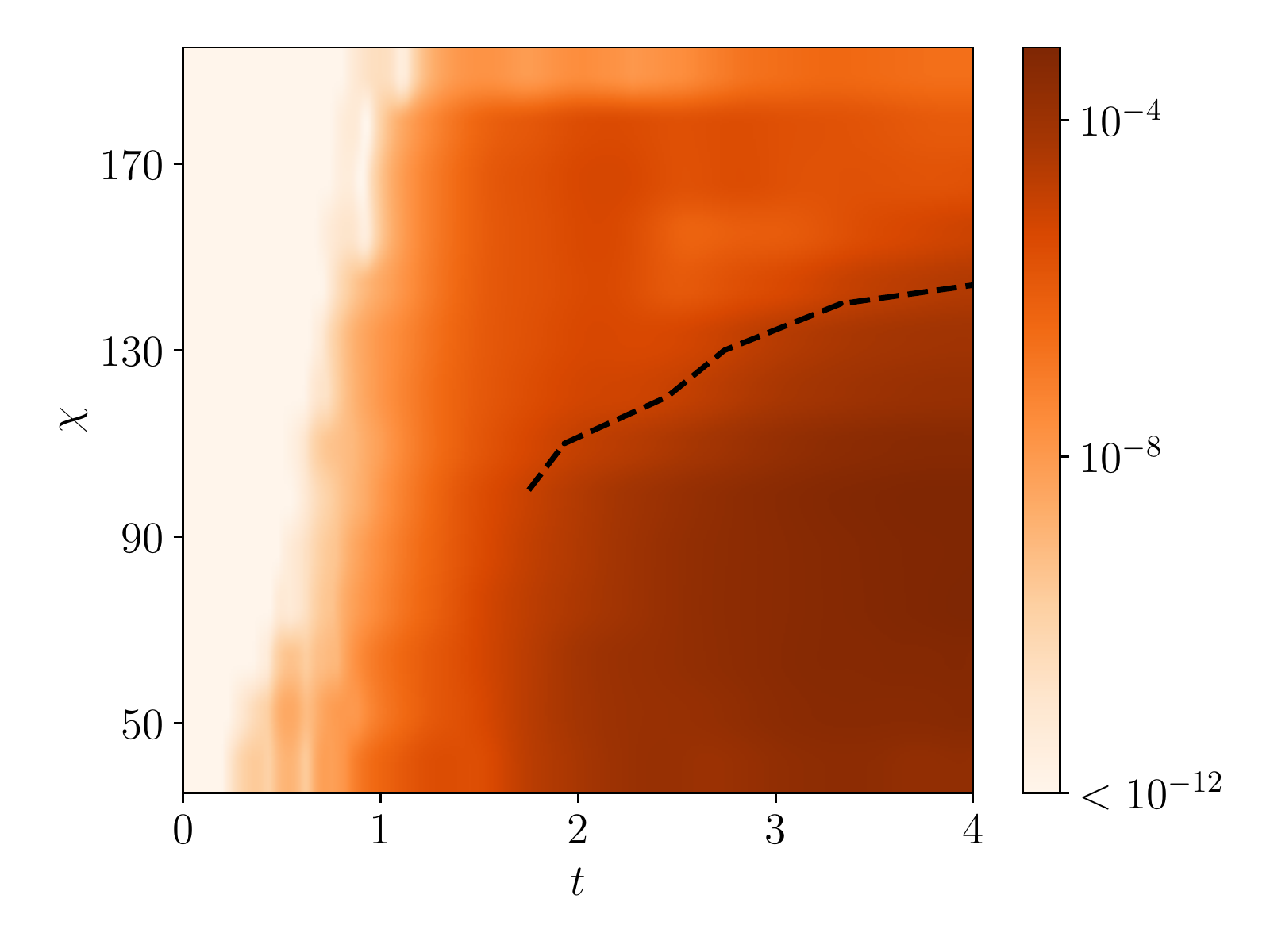}
\centering
  \caption{Error $e(t,\chi)$ of the time-evolved impurity density matrix as a function of bond
  dimension and evolution time (see main text for a precise definition), for an impurity starting from $\rho_{\rm imp}(0)=\ket{\uparrow}\bra{\uparrow}$ and coupled with tunneling amplitude
$t'_{\rm hop}=0.3162$ to a single tight-binding chain of {$L=400$} sites
with homogeneous nearest-neighbor hopping $t_{\rm hop}=1$, initially at
zero temperature and half filling (cf. Ref.~\cite{Nuss15SpatiotemporalFormation}).
  The constant-error $e=10^{-5}$ dashed line   indicates that the required bond
  dimension grows slowly with simulation time. Here we fixed $T=4$, $\delta t=0.01$, $\epsilon=10^{-12}$.}
\label{fig:error}
\end{figure}

{\bf Computational efficiency.}
Finally, we report on the computational efficiency of our method. Previous
works found that for Gaussian ground states~\cite{Fishman2015MPS} and
IFs~\cite{Thoenniss22} (including states with algebraic correlations), the FW
algorithm produces a quantum circuit  of ``local depth'' $D=D(T)$ that scales
at most logarithmically with evolution time $T$. We note that the 
FW  control parameter $\epsilon$ affects the prefactor of $\log T$
scaling of $D$. In turn, $D$ puts an exact upper bound on the bond dimension of
the corresponding MPS as $\chi\le 2^{D}$~\cite{Fishman2015MPS,Thoenniss22},
indicating that computational complexity of the algorithm scales at most
polynomially with evolution time. 

We found that compression of the FW circuit using
conventional singular-value truncation typically leads to a further significant
reduction of the required computational resources. For example, for the data
shown in Fig.~\ref{fig:Cohen_benchmark}, we find a maximum ``local depth'' $D=28$
which sets the hard upper bound $\chi \le 2^{28}$. However, this circuit could be accurately approximated by a MPS with
a much smaller bond dimension $\chi=256=2^8$.

We finally investigated how this MPS compression  affects the \textit{a posteriori} error of  observables. To this end, we considered an environment that consists of a single tight-binding chain~\footnote{The choice of an environment defined by
unitary evolution allows us to avoid errors associated with time-discretization
of a pre-defined spectral density $\Gamma(\omega)$.}. {Having fixed an extremely low FW threshold $\epsilon$ (which makes this source of error negligible),
we estimated
the residual error of time-dependent observables in $t\in[0,T]$ due to the truncated bond dimension,} as the trace distance $e(t,\chi)=\big\lvert \big\lvert \rho_{\rm imp}^{(\chi)}(t)-\rho^{(\infty)}_{\rm imp}(t) \big\rvert\big\rvert_1$ between the reduced density matrix  computed with a cutoff $\chi$ on the IF MPS and the fully converged result (computed using a much higher $\chi=512$). 

The behavior of the
error $e$ as a function of $t$ and $\chi$ is illustrated in Fig.~\ref{fig:error}.
We observe that the bond dimension $\chi=\chi(t)$ required to achieve a fixed error $e$ grows
approximately linearly with $t$, indicating the efficiency of
the approach. We similarly found in other examples we studied, that representing IF
with an MPS with a moderate bond dimension is sufficient to accurately compute
impurity observables. Thus, we conclude that our approach indeed has a polynomial complexity~\cite{lerose2021,Thoenniss22}, allowing one to access long-time impurity dynamics using resources
available in present-day computers.


{\bf Summary and outlook.}
To summarize, we introduced a method for studying dynamics of QIM, based on
a tensor-network representation of reservoir's IF. We applied this approach to
paradigmatic quantum quenches in AIM,  demonstrating that it compares favorably
to state-of-the art 
QMC computations. The approach
is non-perturbative and offers several other advantages: in particular, it
applies to both equilibrium and highly non-equilibrium QIM setups. Moreover,
once a MPS form of the IF is obtained, arbitrary choices of impurity
interactions can be analyzed with modest extra effort. 

We showed that the required computational resources scale polynomially
with the evolution time. Combined with previous results on temporal
entanglement scaling~\cite{Thoenniss22}, this demonstrates that a broad range
of non-equilibrium QIM problems are efficiently solvable using our approach.
While here we focused on quenches of the impurity-reservoir tunnel-coupling in
the single-impurity Anderson model, the approach can be extended to a number of
other setups, including multi-orbital impurities and initial states where
entanglement between impurity and reservoirs is present. Another promising
application is to DMFT, which will require imaginary-time extension of the
technique introduced here. We expect the computational efficiency of the
approach to enable long-time simulations of dynamics in such setups as well,
opening the door to analyzing non-equilibrium behavior of mesoscopic devices
and quantum materials. 

{\bf Acknowledgements.} We thank E. Arrigoni, G. Cohen, M. Eckstein, S. Florens, Y. Ke, M.
Stoudenmire, X. Waintal for discussions. Support by the European Research Council (ERC)
under the European Union’s Horizon 2020 research and innovation programme
(grant agreement No.~864597) and by the Swiss National Science Foundation is
gratefully acknowledged.\\

 While finishing this manuscript, we became aware of a related work by Ng \textit{et al.}, which appeared simultaneously \cite{Ng22Realtime}.

\nocite{wilde2013quantum}
\nocite{footnote_sm_superoperator}
\bibliography{main}

\begin{thebibliography}{66}%
\makeatletter
\providecommand \@ifxundefined [1]{%
 \@ifx{#1\undefined}
}%
\providecommand \@ifnum [1]{%
 \ifnum #1\expandafter \@firstoftwo
 \else \expandafter \@secondoftwo
 \fi
}%
\providecommand \@ifx [1]{%
 \ifx #1\expandafter \@firstoftwo
 \else \expandafter \@secondoftwo
 \fi
}%
\providecommand \natexlab [1]{#1}%
\providecommand \enquote  [1]{``#1''}%
\providecommand \bibnamefont  [1]{#1}%
\providecommand \bibfnamefont [1]{#1}%
\providecommand \citenamefont [1]{#1}%
\providecommand \href@noop [0]{\@secondoftwo}%
\providecommand \href [0]{\begingroup \@sanitize@url \@href}%
\providecommand \@href[1]{\@@startlink{#1}\@@href}%
\providecommand \@@href[1]{\endgroup#1\@@endlink}%
\providecommand \@sanitize@url [0]{\catcode `\\12\catcode `\$12\catcode
  `\&12\catcode `\#12\catcode `\^12\catcode `\_12\catcode `\%12\relax}%
\providecommand \@@startlink[1]{}%
\providecommand \@@endlink[0]{}%
\providecommand \url  [0]{\begingroup\@sanitize@url \@url }%
\providecommand \@url [1]{\endgroup\@href {#1}{\urlprefix }}%
\providecommand \urlprefix  [0]{URL }%
\providecommand \Eprint [0]{\href }%
\providecommand \doibase [0]{https://doi.org/}%
\providecommand \selectlanguage [0]{\@gobble}%
\providecommand \bibinfo  [0]{\@secondoftwo}%
\providecommand \bibfield  [0]{\@secondoftwo}%
\providecommand \translation [1]{[#1]}%
\providecommand \BibitemOpen [0]{}%
\providecommand \bibitemStop [0]{}%
\providecommand \bibitemNoStop [0]{.\EOS\space}%
\providecommand \EOS [0]{\spacefactor3000\relax}%
\providecommand \BibitemShut  [1]{\csname bibitem#1\endcsname}%
\let\auto@bib@innerbib\@empty
\bibitem [{\citenamefont {Bloch}\ \emph {et~al.}(2008)\citenamefont {Bloch},
  \citenamefont {Dalibard},\ and\ \citenamefont {Zwerger}}]{BlochColdAtoms}%
  \BibitemOpen
  \bibfield  {author} {\bibinfo {author} {\bibfnamefont {I.}~\bibnamefont
  {Bloch}}, \bibinfo {author} {\bibfnamefont {J.}~\bibnamefont {Dalibard}},\
  and\ \bibinfo {author} {\bibfnamefont {W.}~\bibnamefont {Zwerger}},\
  }\bibfield  {title} {\bibinfo {title} {Many-body physics with ultracold
  gases},\ }\href {https://doi.org/10.1103/RevModPhys.80.885} {\bibfield
  {journal} {\bibinfo  {journal} {Rev. Mod. Phys.}\ }\textbf {\bibinfo {volume}
  {80}},\ \bibinfo {pages} {885} (\bibinfo {year} {2008})}\BibitemShut
  {NoStop}%
\bibitem [{\citenamefont {Paeckel}\ \emph {et~al.}(2019)\citenamefont
  {Paeckel}, \citenamefont {Köhler}, \citenamefont {Swoboda}, \citenamefont
  {Manmana}, \citenamefont {Schollwöck},\ and\ \citenamefont
  {Hubig}}]{PAECKEL2019167998}%
  \BibitemOpen
  \bibfield  {author} {\bibinfo {author} {\bibfnamefont {S.}~\bibnamefont
  {Paeckel}}, \bibinfo {author} {\bibfnamefont {T.}~\bibnamefont {Köhler}},
  \bibinfo {author} {\bibfnamefont {A.}~\bibnamefont {Swoboda}}, \bibinfo
  {author} {\bibfnamefont {S.~R.}\ \bibnamefont {Manmana}}, \bibinfo {author}
  {\bibfnamefont {U.}~\bibnamefont {Schollwöck}},\ and\ \bibinfo {author}
  {\bibfnamefont {C.}~\bibnamefont {Hubig}},\ }\bibfield  {title} {\bibinfo
  {title} {Time-evolution methods for matrix-product states},\ }\href
  {https://doi.org/https://doi.org/10.1016/j.aop.2019.167998} {\bibfield
  {journal} {\bibinfo  {journal} {Annals of Physics}\ }\textbf {\bibinfo
  {volume} {411}},\ \bibinfo {pages} {167998} (\bibinfo {year}
  {2019})}\BibitemShut {NoStop}%
\bibitem [{\citenamefont {Aoki}\ \emph {et~al.}(2014)\citenamefont {Aoki},
  \citenamefont {Tsuji}, \citenamefont {Eckstein}, \citenamefont {Kollar},
  \citenamefont {Oka},\ and\ \citenamefont {Werner}}]{aoki14neqdmftreview}%
  \BibitemOpen
  \bibfield  {author} {\bibinfo {author} {\bibfnamefont {H.}~\bibnamefont
  {Aoki}}, \bibinfo {author} {\bibfnamefont {N.}~\bibnamefont {Tsuji}},
  \bibinfo {author} {\bibfnamefont {M.}~\bibnamefont {Eckstein}}, \bibinfo
  {author} {\bibfnamefont {M.}~\bibnamefont {Kollar}}, \bibinfo {author}
  {\bibfnamefont {T.}~\bibnamefont {Oka}},\ and\ \bibinfo {author}
  {\bibfnamefont {P.}~\bibnamefont {Werner}},\ }\bibfield  {title} {\bibinfo
  {title} {Nonequilibrium dynamical mean-field theory and its applications},\
  }\href {https://doi.org/10.1103/RevModPhys.86.779} {\bibfield  {journal}
  {\bibinfo  {journal} {Rev. Mod. Phys.}\ }\textbf {\bibinfo {volume} {86}},\
  \bibinfo {pages} {779} (\bibinfo {year} {2014})}\BibitemShut {NoStop}%
\bibitem [{\citenamefont {{Pustilnik}}\ and\ \citenamefont
  {{Glazman}}(2004)}]{GlazmanKondo2004}%
  \BibitemOpen
  \bibfield  {author} {\bibinfo {author} {\bibfnamefont {M.}~\bibnamefont
  {{Pustilnik}}}\ and\ \bibinfo {author} {\bibfnamefont {L.}~\bibnamefont
  {{Glazman}}},\ }\bibfield  {title} {\bibinfo {title} {{TOPICAL REVIEW: Kondo
  effect in quantum dots}},\ }\href
  {https://doi.org/10.1088/0953-8984/16/16/R01} {\bibfield  {journal} {\bibinfo
   {journal} {Journal of Physics Condensed Matter}\ }\textbf {\bibinfo {volume}
  {16}},\ \bibinfo {pages} {R513} (\bibinfo {year} {2004})},\ \Eprint
  {https://arxiv.org/abs/cond-mat/0401517} {arXiv:cond-mat/0401517
  [cond-mat.mes-hall]} \BibitemShut {NoStop}%
\bibitem [{\citenamefont {Kan\'asz-Nagy}\ \emph {et~al.}(2018)\citenamefont
  {Kan\'asz-Nagy}, \citenamefont {Ashida}, \citenamefont {Shi}, \citenamefont
  {Moca}, \citenamefont {Ikeda}, \citenamefont {F\"olling}, \citenamefont
  {Cirac}, \citenamefont {Zar\'and},\ and\ \citenamefont
  {Demler}}]{KanaszNagy18ExploringKondo}%
  \BibitemOpen
  \bibfield  {author} {\bibinfo {author} {\bibfnamefont {M.}~\bibnamefont
  {Kan\'asz-Nagy}}, \bibinfo {author} {\bibfnamefont {Y.}~\bibnamefont
  {Ashida}}, \bibinfo {author} {\bibfnamefont {T.}~\bibnamefont {Shi}},
  \bibinfo {author} {\bibfnamefont {C.}~\bibnamefont {Moca}}, \bibinfo {author}
  {\bibfnamefont {T.~N.}\ \bibnamefont {Ikeda}}, \bibinfo {author}
  {\bibfnamefont {S.}~\bibnamefont {F\"olling}}, \bibinfo {author}
  {\bibfnamefont {J.~I.}\ \bibnamefont {Cirac}}, \bibinfo {author}
  {\bibfnamefont {G.}~\bibnamefont {Zar\'and}},\ and\ \bibinfo {author}
  {\bibfnamefont {E.~A.}\ \bibnamefont {Demler}},\ }\bibfield  {title}
  {\bibinfo {title} {Exploring the anisotropic kondo model in and out of
  equilibrium with alkaline-earth atoms},\ }\href
  {https://doi.org/10.1103/PhysRevB.97.155156} {\bibfield  {journal} {\bibinfo
  {journal} {Phys. Rev. B}\ }\textbf {\bibinfo {volume} {97}},\ \bibinfo
  {pages} {155156} (\bibinfo {year} {2018})}\BibitemShut {NoStop}%
\bibitem [{\citenamefont {Riegger}\ \emph {et~al.}(2018)\citenamefont
  {Riegger}, \citenamefont {Darkwah~Oppong}, \citenamefont {H\"ofer},
  \citenamefont {Fernandes}, \citenamefont {Bloch},\ and\ \citenamefont
  {F\"olling}}]{Riegger18LocalizedMagneticMoments}%
  \BibitemOpen
  \bibfield  {author} {\bibinfo {author} {\bibfnamefont {L.}~\bibnamefont
  {Riegger}}, \bibinfo {author} {\bibfnamefont {N.}~\bibnamefont
  {Darkwah~Oppong}}, \bibinfo {author} {\bibfnamefont {M.}~\bibnamefont
  {H\"ofer}}, \bibinfo {author} {\bibfnamefont {D.~R.}\ \bibnamefont
  {Fernandes}}, \bibinfo {author} {\bibfnamefont {I.}~\bibnamefont {Bloch}},\
  and\ \bibinfo {author} {\bibfnamefont {S.}~\bibnamefont {F\"olling}},\
  }\bibfield  {title} {\bibinfo {title} {Localized magnetic moments with
  tunable spin exchange in a gas of ultracold fermions},\ }\href
  {https://doi.org/10.1103/PhysRevLett.120.143601} {\bibfield  {journal}
  {\bibinfo  {journal} {Phys. Rev. Lett.}\ }\textbf {\bibinfo {volume} {120}},\
  \bibinfo {pages} {143601} (\bibinfo {year} {2018})}\BibitemShut {NoStop}%
\bibitem [{\citenamefont {Anderson}(1961)}]{AIM}%
  \BibitemOpen
  \bibfield  {author} {\bibinfo {author} {\bibfnamefont {P.~W.}\ \bibnamefont
  {Anderson}},\ }\bibfield  {title} {\bibinfo {title} {Localized magnetic
  states in metals},\ }\href {https://doi.org/10.1103/PhysRev.124.41}
  {\bibfield  {journal} {\bibinfo  {journal} {Phys. Rev.}\ }\textbf {\bibinfo
  {volume} {124}},\ \bibinfo {pages} {41} (\bibinfo {year} {1961})}\BibitemShut
  {NoStop}%
\bibitem [{\citenamefont {Hewson}(1993)}]{hewson_1993}%
  \BibitemOpen
  \bibfield  {author} {\bibinfo {author} {\bibfnamefont {A.~C.}\ \bibnamefont
  {Hewson}},\ }\href {https://doi.org/10.1017/CBO9780511470752} {\emph
  {\bibinfo {title} {The Kondo Problem to Heavy Fermions}}},\ Cambridge Studies
  in Magnetism\ (\bibinfo  {publisher} {Cambridge University Press},\ \bibinfo
  {year} {1993})\BibitemShut {NoStop}%
\bibitem [{\citenamefont {Georges}\ \emph {et~al.}(1996)\citenamefont
  {Georges}, \citenamefont {Kotliar}, \citenamefont {Krauth},\ and\
  \citenamefont {Rozenberg}}]{GeorgesRMP}%
  \BibitemOpen
  \bibfield  {author} {\bibinfo {author} {\bibfnamefont {A.}~\bibnamefont
  {Georges}}, \bibinfo {author} {\bibfnamefont {G.}~\bibnamefont {Kotliar}},
  \bibinfo {author} {\bibfnamefont {W.}~\bibnamefont {Krauth}},\ and\ \bibinfo
  {author} {\bibfnamefont {M.~J.}\ \bibnamefont {Rozenberg}},\ }\bibfield
  {title} {\bibinfo {title} {Dynamical mean-field theory of strongly correlated
  fermion systems and the limit of infinite dimensions},\ }\href
  {https://doi.org/10.1103/RevModPhys.68.13} {\bibfield  {journal} {\bibinfo
  {journal} {Rev. Mod. Phys.}\ }\textbf {\bibinfo {volume} {68}},\ \bibinfo
  {pages} {13} (\bibinfo {year} {1996})}\BibitemShut {NoStop}%
\bibitem [{\citenamefont {{Makarov}}\ and\ \citenamefont
  {{Makri}}(1994)}]{MakriMakarov94}%
  \BibitemOpen
  \bibfield  {author} {\bibinfo {author} {\bibfnamefont {D.~E.}\ \bibnamefont
  {{Makarov}}}\ and\ \bibinfo {author} {\bibfnamefont {N.}~\bibnamefont
  {{Makri}}},\ }\bibfield  {title} {\bibinfo {title} {{Path integrals for
  dissipative systems by tensor multiplication. Condensed phase quantum
  dynamics for arbitrarily long time}},\ }\href
  {https://doi.org/10.1016/0009-2614(94)00275-4} {\bibfield  {journal}
  {\bibinfo  {journal} {Chemical Physics Letters}\ }\textbf {\bibinfo {volume}
  {221}},\ \bibinfo {pages} {482} (\bibinfo {year} {1994})}\BibitemShut
  {NoStop}%
\bibitem [{\citenamefont {Weiss}\ \emph {et~al.}(2008)\citenamefont {Weiss},
  \citenamefont {Eckel}, \citenamefont {Thorwart},\ and\ \citenamefont
  {Egger}}]{EggerIterative2008}%
  \BibitemOpen
  \bibfield  {author} {\bibinfo {author} {\bibfnamefont {S.}~\bibnamefont
  {Weiss}}, \bibinfo {author} {\bibfnamefont {J.}~\bibnamefont {Eckel}},
  \bibinfo {author} {\bibfnamefont {M.}~\bibnamefont {Thorwart}},\ and\
  \bibinfo {author} {\bibfnamefont {R.}~\bibnamefont {Egger}},\ }\bibfield
  {title} {\bibinfo {title} {Iterative real-time path integral approach to
  nonequilibrium quantum transport},\ }\href
  {https://doi.org/10.1103/PhysRevB.77.195316} {\bibfield  {journal} {\bibinfo
  {journal} {Phys. Rev. B}\ }\textbf {\bibinfo {volume} {77}},\ \bibinfo
  {pages} {195316} (\bibinfo {year} {2008})}\BibitemShut {NoStop}%
\bibitem [{\citenamefont {Segal}\ \emph {et~al.}(2010)\citenamefont {Segal},
  \citenamefont {Millis},\ and\ \citenamefont {Reichman}}]{MillisImp2010}%
  \BibitemOpen
  \bibfield  {author} {\bibinfo {author} {\bibfnamefont {D.}~\bibnamefont
  {Segal}}, \bibinfo {author} {\bibfnamefont {A.~J.}\ \bibnamefont {Millis}},\
  and\ \bibinfo {author} {\bibfnamefont {D.~R.}\ \bibnamefont {Reichman}},\
  }\bibfield  {title} {\bibinfo {title} {Numerically exact path-integral
  simulation of nonequilibrium quantum transport and dissipation},\ }\href
  {https://doi.org/10.1103/PhysRevB.82.205323} {\bibfield  {journal} {\bibinfo
  {journal} {Phys. Rev. B}\ }\textbf {\bibinfo {volume} {82}},\ \bibinfo
  {pages} {205323} (\bibinfo {year} {2010})}\BibitemShut {NoStop}%
\bibitem [{\citenamefont {Tu}\ and\ \citenamefont
  {Zhang}(2008)}]{Tu08nonmarkovian}%
  \BibitemOpen
  \bibfield  {author} {\bibinfo {author} {\bibfnamefont {M.~W.~Y.}\
  \bibnamefont {Tu}}\ and\ \bibinfo {author} {\bibfnamefont {W.-M.}\
  \bibnamefont {Zhang}},\ }\bibfield  {title} {\bibinfo {title} {Non-markovian
  decoherence theory for a double-dot charge qubit},\ }\href
  {https://doi.org/10.1103/PhysRevB.78.235311} {\bibfield  {journal} {\bibinfo
  {journal} {Phys. Rev. B}\ }\textbf {\bibinfo {volume} {78}},\ \bibinfo
  {pages} {235311} (\bibinfo {year} {2008})}\BibitemShut {NoStop}%
\bibitem [{\citenamefont {{Jin}}\ \emph {et~al.}(2010)\citenamefont {{Jin}},
  \citenamefont {{Wei-Yuan Tu}}, \citenamefont {{Zhang}},\ and\ \citenamefont
  {{Yan}}}]{IFnanodevices}%
  \BibitemOpen
  \bibfield  {author} {\bibinfo {author} {\bibfnamefont {J.}~\bibnamefont
  {{Jin}}}, \bibinfo {author} {\bibfnamefont {M.}~\bibnamefont {{Wei-Yuan
  Tu}}}, \bibinfo {author} {\bibfnamefont {W.-M.}\ \bibnamefont {{Zhang}}},\
  and\ \bibinfo {author} {\bibfnamefont {Y.}~\bibnamefont {{Yan}}},\ }\bibfield
   {title} {\bibinfo {title} {{Non-equilibrium quantum theory for nanodevices
  based on the Feynman-Vernon influence functional}},\ }\href
  {https://doi.org/10.1088/1367-2630/12/8/083013} {\bibfield  {journal}
  {\bibinfo  {journal} {New Journal of Physics}\ }\textbf {\bibinfo {volume}
  {12}},\ \bibinfo {eid} {083013} (\bibinfo {year} {2010})}\BibitemShut
  {NoStop}%
\bibitem [{\citenamefont {Dorda}\ \emph {et~al.}(2014)\citenamefont {Dorda},
  \citenamefont {Nuss}, \citenamefont {von~der Linden},\ and\ \citenamefont
  {Arrigoni}}]{dorda14auxiliary}%
  \BibitemOpen
  \bibfield  {author} {\bibinfo {author} {\bibfnamefont {A.}~\bibnamefont
  {Dorda}}, \bibinfo {author} {\bibfnamefont {M.}~\bibnamefont {Nuss}},
  \bibinfo {author} {\bibfnamefont {W.}~\bibnamefont {von~der Linden}},\ and\
  \bibinfo {author} {\bibfnamefont {E.}~\bibnamefont {Arrigoni}},\ }\bibfield
  {title} {\bibinfo {title} {Auxiliary master equation approach to
  nonequilibrium correlated impurities},\ }\href
  {https://doi.org/10.1103/PhysRevB.89.165105} {\bibfield  {journal} {\bibinfo
  {journal} {Phys. Rev. B}\ }\textbf {\bibinfo {volume} {89}},\ \bibinfo
  {pages} {165105} (\bibinfo {year} {2014})}\BibitemShut {NoStop}%
\bibitem [{\citenamefont {Lotem}\ \emph {et~al.}(2020)\citenamefont {Lotem},
  \citenamefont {Weichselbaum}, \citenamefont {von Delft},\ and\ \citenamefont
  {Goldstein}}]{Lotem20renormalized}%
  \BibitemOpen
  \bibfield  {author} {\bibinfo {author} {\bibfnamefont {M.}~\bibnamefont
  {Lotem}}, \bibinfo {author} {\bibfnamefont {A.}~\bibnamefont {Weichselbaum}},
  \bibinfo {author} {\bibfnamefont {J.}~\bibnamefont {von Delft}},\ and\
  \bibinfo {author} {\bibfnamefont {M.}~\bibnamefont {Goldstein}},\ }\bibfield
  {title} {\bibinfo {title} {Renormalized lindblad driving: A numerically exact
  nonequilibrium quantum impurity solver},\ }\href
  {https://doi.org/10.1103/PhysRevResearch.2.043052} {\bibfield  {journal}
  {\bibinfo  {journal} {Phys. Rev. Research}\ }\textbf {\bibinfo {volume}
  {2}},\ \bibinfo {pages} {043052} (\bibinfo {year} {2020})}\BibitemShut
  {NoStop}%
\bibitem [{\citenamefont {Tanimura}\ and\ \citenamefont
  {Kubo}(1989)}]{Tanimura89}%
  \BibitemOpen
  \bibfield  {author} {\bibinfo {author} {\bibfnamefont {Y.}~\bibnamefont
  {Tanimura}}\ and\ \bibinfo {author} {\bibfnamefont {R.}~\bibnamefont
  {Kubo}},\ }\bibfield  {title} {\bibinfo {title} {Time evolution of a quantum
  system in contact with a nearly gaussian-markoffian noise bath},\ }\href
  {https://doi.org/10.1143/JPSJ.58.101} {\bibfield  {journal} {\bibinfo
  {journal} {Journal of the Physical Society of Japan}\ }\textbf {\bibinfo
  {volume} {58}},\ \bibinfo {pages} {101} (\bibinfo {year} {1989})},\ \Eprint
  {https://arxiv.org/abs/https://doi.org/10.1143/JPSJ.58.101}
  {https://doi.org/10.1143/JPSJ.58.101} \BibitemShut {NoStop}%
\bibitem [{\citenamefont {Jin}\ \emph {et~al.}(2008)\citenamefont {Jin},
  \citenamefont {Zheng},\ and\ \citenamefont {Yan}}]{Jin08heom}%
  \BibitemOpen
  \bibfield  {author} {\bibinfo {author} {\bibfnamefont {J.}~\bibnamefont
  {Jin}}, \bibinfo {author} {\bibfnamefont {X.}~\bibnamefont {Zheng}},\ and\
  \bibinfo {author} {\bibfnamefont {Y.}~\bibnamefont {Yan}},\ }\bibfield
  {title} {\bibinfo {title} {Exact dynamics of dissipative electronic systems
  and quantum transport: Hierarchical equations of motion approach},\ }\href
  {https://doi.org/10.1063/1.2938087} {\bibfield  {journal} {\bibinfo
  {journal} {The Journal of Chemical Physics}\ }\textbf {\bibinfo {volume}
  {128}},\ \bibinfo {pages} {234703} (\bibinfo {year} {2008})},\ \Eprint
  {https://arxiv.org/abs/https://doi.org/10.1063/1.2938087}
  {https://doi.org/10.1063/1.2938087} \BibitemShut {NoStop}%
\bibitem [{\citenamefont {Dan}\ \emph {et~al.}(2022)\citenamefont {Dan},
  \citenamefont {Xu}, \citenamefont {Ankerhold},\ and\ \citenamefont
  {Shi}}]{Dan22efficientheom}%
  \BibitemOpen
  \bibfield  {author} {\bibinfo {author} {\bibfnamefont {X.}~\bibnamefont
  {Dan}}, \bibinfo {author} {\bibfnamefont {J.~T.}\ \bibnamefont {Xu},
  \bibfnamefont {Meng~Stockburger}}, \bibinfo {author} {\bibfnamefont
  {J.}~\bibnamefont {Ankerhold}},\ and\ \bibinfo {author} {\bibfnamefont
  {Q.}~\bibnamefont {Shi}},\ }\bibfield  {title} {\bibinfo {title} {Efficient
  low temperature simulations for fermionic reservoirs with the hierarchical
  equations of motion method: Application to the anderson impurity model}\
  }\href {https://doi.org/10.48550/arXiv.2211.04089}
  {10.48550/arXiv.2211.04089} (\bibinfo {year} {2022})\BibitemShut {NoStop}%
\bibitem [{\citenamefont {Anders}\ and\ \citenamefont
  {Schiller}(2005)}]{Anders05Realtime}%
  \BibitemOpen
  \bibfield  {author} {\bibinfo {author} {\bibfnamefont {F.~B.}\ \bibnamefont
  {Anders}}\ and\ \bibinfo {author} {\bibfnamefont {A.}~\bibnamefont
  {Schiller}},\ }\bibfield  {title} {\bibinfo {title} {Real-time dynamics in
  quantum-impurity systems: A time-dependent numerical renormalization-group
  approach},\ }\href {https://doi.org/10.1103/PhysRevLett.95.196801} {\bibfield
   {journal} {\bibinfo  {journal} {Phys. Rev. Lett.}\ }\textbf {\bibinfo
  {volume} {95}},\ \bibinfo {pages} {196801} (\bibinfo {year}
  {2005})}\BibitemShut {NoStop}%
\bibitem [{\citenamefont {Nghiem}\ and\ \citenamefont
  {Costi}(2017)}]{nghiem17timeevolution}%
  \BibitemOpen
  \bibfield  {author} {\bibinfo {author} {\bibfnamefont {H.~T.~M.}\
  \bibnamefont {Nghiem}}\ and\ \bibinfo {author} {\bibfnamefont {T.~A.}\
  \bibnamefont {Costi}},\ }\bibfield  {title} {\bibinfo {title} {Time evolution
  of the kondo resonance in response to a quench},\ }\href
  {https://doi.org/10.1103/PhysRevLett.119.156601} {\bibfield  {journal}
  {\bibinfo  {journal} {Phys. Rev. Lett.}\ }\textbf {\bibinfo {volume} {119}},\
  \bibinfo {pages} {156601} (\bibinfo {year} {2017})}\BibitemShut {NoStop}%
\bibitem [{\citenamefont {Schwarz}\ \emph {et~al.}(2018)\citenamefont
  {Schwarz}, \citenamefont {Weymann}, \citenamefont {von Delft},\ and\
  \citenamefont {Weichselbaum}}]{DelftNonEquil18}%
  \BibitemOpen
  \bibfield  {author} {\bibinfo {author} {\bibfnamefont {F.}~\bibnamefont
  {Schwarz}}, \bibinfo {author} {\bibfnamefont {I.}~\bibnamefont {Weymann}},
  \bibinfo {author} {\bibfnamefont {J.}~\bibnamefont {von Delft}},\ and\
  \bibinfo {author} {\bibfnamefont {A.}~\bibnamefont {Weichselbaum}},\
  }\bibfield  {title} {\bibinfo {title} {Nonequilibrium steady-state transport
  in quantum impurity models: A thermofield and quantum quench approach using
  matrix product states},\ }\href
  {https://doi.org/10.1103/PhysRevLett.121.137702} {\bibfield  {journal}
  {\bibinfo  {journal} {Phys. Rev. Lett.}\ }\textbf {\bibinfo {volume} {121}},\
  \bibinfo {pages} {137702} (\bibinfo {year} {2018})}\BibitemShut {NoStop}%
\bibitem [{\citenamefont {Prior}\ \emph {et~al.}(2010)\citenamefont {Prior},
  \citenamefont {Chin}, \citenamefont {Huelga},\ and\ \citenamefont
  {Plenio}}]{prior10efficient}%
  \BibitemOpen
  \bibfield  {author} {\bibinfo {author} {\bibfnamefont {J.}~\bibnamefont
  {Prior}}, \bibinfo {author} {\bibfnamefont {A.~W.}\ \bibnamefont {Chin}},
  \bibinfo {author} {\bibfnamefont {S.~F.}\ \bibnamefont {Huelga}},\ and\
  \bibinfo {author} {\bibfnamefont {M.~B.}\ \bibnamefont {Plenio}},\ }\bibfield
   {title} {\bibinfo {title} {Efficient simulation of strong system-environment
  interactions},\ }\href {https://doi.org/10.1103/PhysRevLett.105.050404}
  {\bibfield  {journal} {\bibinfo  {journal} {Phys. Rev. Lett.}\ }\textbf
  {\bibinfo {volume} {105}},\ \bibinfo {pages} {050404} (\bibinfo {year}
  {2010})}\BibitemShut {NoStop}%
\bibitem [{\citenamefont {Nuss}\ \emph {et~al.}(2015)\citenamefont {Nuss},
  \citenamefont {Ganahl}, \citenamefont {Arrigoni}, \citenamefont {von~der
  Linden},\ and\ \citenamefont {Evertz}}]{Nuss15SpatiotemporalFormation}%
  \BibitemOpen
  \bibfield  {author} {\bibinfo {author} {\bibfnamefont {M.}~\bibnamefont
  {Nuss}}, \bibinfo {author} {\bibfnamefont {M.}~\bibnamefont {Ganahl}},
  \bibinfo {author} {\bibfnamefont {E.}~\bibnamefont {Arrigoni}}, \bibinfo
  {author} {\bibfnamefont {W.}~\bibnamefont {von~der Linden}},\ and\ \bibinfo
  {author} {\bibfnamefont {H.~G.}\ \bibnamefont {Evertz}},\ }\bibfield  {title}
  {\bibinfo {title} {Nonequilibrium spatiotemporal formation of the kondo
  screening cloud on a lattice},\ }\href
  {https://doi.org/10.1103/PhysRevB.91.085127} {\bibfield  {journal} {\bibinfo
  {journal} {Phys. Rev. B}\ }\textbf {\bibinfo {volume} {91}},\ \bibinfo
  {pages} {085127} (\bibinfo {year} {2015})}\BibitemShut {NoStop}%
\bibitem [{\citenamefont {Wolf}\ \emph {et~al.}(2014)\citenamefont {Wolf},
  \citenamefont {McCulloch},\ and\ \citenamefont {Schollw\"ock}}]{WolfPRB14}%
  \BibitemOpen
  \bibfield  {author} {\bibinfo {author} {\bibfnamefont {F.~A.}\ \bibnamefont
  {Wolf}}, \bibinfo {author} {\bibfnamefont {I.~P.}\ \bibnamefont
  {McCulloch}},\ and\ \bibinfo {author} {\bibfnamefont {U.}~\bibnamefont
  {Schollw\"ock}},\ }\bibfield  {title} {\bibinfo {title} {Solving
  nonequilibrium dynamical mean-field theory using matrix product states},\
  }\href {https://doi.org/10.1103/PhysRevB.90.235131} {\bibfield  {journal}
  {\bibinfo  {journal} {Phys. Rev. B}\ }\textbf {\bibinfo {volume} {90}},\
  \bibinfo {pages} {235131} (\bibinfo {year} {2014})}\BibitemShut {NoStop}%
\bibitem [{\citenamefont {N\"u\ss{}eler}\ \emph {et~al.}(2020)\citenamefont
  {N\"u\ss{}eler}, \citenamefont {Dhand}, \citenamefont {Huelga},\ and\
  \citenamefont {Plenio}}]{Nusseler20Efficient}%
  \BibitemOpen
  \bibfield  {author} {\bibinfo {author} {\bibfnamefont {A.}~\bibnamefont
  {N\"u\ss{}eler}}, \bibinfo {author} {\bibfnamefont {I.}~\bibnamefont
  {Dhand}}, \bibinfo {author} {\bibfnamefont {S.~F.}\ \bibnamefont {Huelga}},\
  and\ \bibinfo {author} {\bibfnamefont {M.~B.}\ \bibnamefont {Plenio}},\
  }\bibfield  {title} {\bibinfo {title} {Efficient simulation of open quantum
  systems coupled to a fermionic bath},\ }\href
  {https://doi.org/10.1103/PhysRevB.101.155134} {\bibfield  {journal} {\bibinfo
   {journal} {Phys. Rev. B}\ }\textbf {\bibinfo {volume} {101}},\ \bibinfo
  {pages} {155134} (\bibinfo {year} {2020})}\BibitemShut {NoStop}%
\bibitem [{\citenamefont {W\'ojtowicz}\ \emph {et~al.}(2020)\citenamefont
  {W\'ojtowicz}, \citenamefont {Elenewski}, \citenamefont {Rams},\ and\
  \citenamefont {Zwolak}}]{Wojtowicz20OpensystemTN}%
  \BibitemOpen
  \bibfield  {author} {\bibinfo {author} {\bibfnamefont {G.}~\bibnamefont
  {W\'ojtowicz}}, \bibinfo {author} {\bibfnamefont {J.~E.}\ \bibnamefont
  {Elenewski}}, \bibinfo {author} {\bibfnamefont {M.~M.}\ \bibnamefont
  {Rams}},\ and\ \bibinfo {author} {\bibfnamefont {M.}~\bibnamefont {Zwolak}},\
  }\bibfield  {title} {\bibinfo {title} {Open-system tensor networks and
  kramers' crossover for quantum transport},\ }\href
  {https://doi.org/10.1103/PhysRevA.101.050301} {\bibfield  {journal} {\bibinfo
   {journal} {Phys. Rev. A}\ }\textbf {\bibinfo {volume} {101}},\ \bibinfo
  {pages} {050301} (\bibinfo {year} {2020})}\BibitemShut {NoStop}%
\bibitem [{\citenamefont {Kohn}\ and\ \citenamefont
  {Santoro}(2022)}]{Kohn_2022}%
  \BibitemOpen
  \bibfield  {author} {\bibinfo {author} {\bibfnamefont {L.}~\bibnamefont
  {Kohn}}\ and\ \bibinfo {author} {\bibfnamefont {G.~E.}\ \bibnamefont
  {Santoro}},\ }\bibfield  {title} {\bibinfo {title} {Quench dynamics of the
  anderson impurity model at finite temperature using matrix product states:
  entanglement and bath dynamics},\ }\href
  {https://doi.org/10.1088/1742-5468/ac729b} {\bibfield  {journal} {\bibinfo
  {journal} {Journal of Statistical Mechanics: Theory and Experiment}\ }\textbf
  {\bibinfo {volume} {2022}},\ \bibinfo {pages} {063102} (\bibinfo {year}
  {2022})}\BibitemShut {NoStop}%
\bibitem [{\citenamefont {M\"uhlbacher}\ and\ \citenamefont
  {Rabani}(2008)}]{Muhlbacher08Realtime}%
  \BibitemOpen
  \bibfield  {author} {\bibinfo {author} {\bibfnamefont {L.}~\bibnamefont
  {M\"uhlbacher}}\ and\ \bibinfo {author} {\bibfnamefont {E.}~\bibnamefont
  {Rabani}},\ }\bibfield  {title} {\bibinfo {title} {Real-time path integral
  approach to nonequilibrium many-body quantum systems},\ }\href
  {https://doi.org/10.1103/PhysRevLett.100.176403} {\bibfield  {journal}
  {\bibinfo  {journal} {Phys. Rev. Lett.}\ }\textbf {\bibinfo {volume} {100}},\
  \bibinfo {pages} {176403} (\bibinfo {year} {2008})}\BibitemShut {NoStop}%
\bibitem [{\citenamefont {Schir\'o}\ and\ \citenamefont
  {Fabrizio}(2009)}]{Schiro09realtime}%
  \BibitemOpen
  \bibfield  {author} {\bibinfo {author} {\bibfnamefont {M.}~\bibnamefont
  {Schir\'o}}\ and\ \bibinfo {author} {\bibfnamefont {M.}~\bibnamefont
  {Fabrizio}},\ }\bibfield  {title} {\bibinfo {title} {Real-time diagrammatic
  monte carlo for nonequilibrium quantum transport},\ }\href
  {https://doi.org/10.1103/PhysRevB.79.153302} {\bibfield  {journal} {\bibinfo
  {journal} {Phys. Rev. B}\ }\textbf {\bibinfo {volume} {79}},\ \bibinfo
  {pages} {153302} (\bibinfo {year} {2009})}\BibitemShut {NoStop}%
\bibitem [{\citenamefont {Werner}\ \emph {et~al.}(2009)\citenamefont {Werner},
  \citenamefont {Oka},\ and\ \citenamefont {Millis}}]{Werner09Diagrammatic}%
  \BibitemOpen
  \bibfield  {author} {\bibinfo {author} {\bibfnamefont {P.}~\bibnamefont
  {Werner}}, \bibinfo {author} {\bibfnamefont {T.}~\bibnamefont {Oka}},\ and\
  \bibinfo {author} {\bibfnamefont {A.~J.}\ \bibnamefont {Millis}},\ }\bibfield
   {title} {\bibinfo {title} {Diagrammatic monte carlo simulation of
  nonequilibrium systems},\ }\href {https://doi.org/10.1103/PhysRevB.79.035320}
  {\bibfield  {journal} {\bibinfo  {journal} {Phys. Rev. B}\ }\textbf {\bibinfo
  {volume} {79}},\ \bibinfo {pages} {035320} (\bibinfo {year}
  {2009})}\BibitemShut {NoStop}%
\bibitem [{\citenamefont {Gull}\ \emph {et~al.}(2011)\citenamefont {Gull},
  \citenamefont {Reichman},\ and\ \citenamefont
  {Millis}}]{gull11NumericallyExact}%
  \BibitemOpen
  \bibfield  {author} {\bibinfo {author} {\bibfnamefont {E.}~\bibnamefont
  {Gull}}, \bibinfo {author} {\bibfnamefont {D.~R.}\ \bibnamefont {Reichman}},\
  and\ \bibinfo {author} {\bibfnamefont {A.~J.}\ \bibnamefont {Millis}},\
  }\bibfield  {title} {\bibinfo {title} {Numerically exact long-time behavior
  of nonequilibrium quantum impurity models},\ }\href
  {https://doi.org/10.1103/PhysRevB.84.085134} {\bibfield  {journal} {\bibinfo
  {journal} {Phys. Rev. B}\ }\textbf {\bibinfo {volume} {84}},\ \bibinfo
  {pages} {085134} (\bibinfo {year} {2011})}\BibitemShut {NoStop}%
\bibitem [{\citenamefont {Cohen}\ and\ \citenamefont
  {Rabani}(2011)}]{cohen11memory}%
  \BibitemOpen
  \bibfield  {author} {\bibinfo {author} {\bibfnamefont {G.}~\bibnamefont
  {Cohen}}\ and\ \bibinfo {author} {\bibfnamefont {E.}~\bibnamefont {Rabani}},\
  }\bibfield  {title} {\bibinfo {title} {Memory effects in nonequilibrium
  quantum impurity models},\ }\href
  {https://doi.org/10.1103/PhysRevB.84.075150} {\bibfield  {journal} {\bibinfo
  {journal} {Phys. Rev. B}\ }\textbf {\bibinfo {volume} {84}},\ \bibinfo
  {pages} {075150} (\bibinfo {year} {2011})}\BibitemShut {NoStop}%
\bibitem [{\citenamefont {Cohen}\ \emph {et~al.}(2013)\citenamefont {Cohen},
  \citenamefont {Gull}, \citenamefont {Reichman}, \citenamefont {Millis},\ and\
  \citenamefont {Rabani}}]{cohen13neqkondo}%
  \BibitemOpen
  \bibfield  {author} {\bibinfo {author} {\bibfnamefont {G.}~\bibnamefont
  {Cohen}}, \bibinfo {author} {\bibfnamefont {E.}~\bibnamefont {Gull}},
  \bibinfo {author} {\bibfnamefont {D.~R.}\ \bibnamefont {Reichman}}, \bibinfo
  {author} {\bibfnamefont {A.~J.}\ \bibnamefont {Millis}},\ and\ \bibinfo
  {author} {\bibfnamefont {E.}~\bibnamefont {Rabani}},\ }\bibfield  {title}
  {\bibinfo {title} {Numerically exact long-time magnetization dynamics at the
  nonequilibrium kondo crossover of the anderson impurity model},\ }\href
  {https://doi.org/10.1103/PhysRevB.87.195108} {\bibfield  {journal} {\bibinfo
  {journal} {Phys. Rev. B}\ }\textbf {\bibinfo {volume} {87}},\ \bibinfo
  {pages} {195108} (\bibinfo {year} {2013})}\BibitemShut {NoStop}%
\bibitem [{\citenamefont {Ashida}\ \emph {et~al.}(2018)\citenamefont {Ashida},
  \citenamefont {Shi}, \citenamefont {Ba\~nuls}, \citenamefont {Cirac},\ and\
  \citenamefont {Demler}}]{Ashida2018}%
  \BibitemOpen
  \bibfield  {author} {\bibinfo {author} {\bibfnamefont {Y.}~\bibnamefont
  {Ashida}}, \bibinfo {author} {\bibfnamefont {T.}~\bibnamefont {Shi}},
  \bibinfo {author} {\bibfnamefont {M.~C.}\ \bibnamefont {Ba\~nuls}}, \bibinfo
  {author} {\bibfnamefont {J.~I.}\ \bibnamefont {Cirac}},\ and\ \bibinfo
  {author} {\bibfnamefont {E.}~\bibnamefont {Demler}},\ }\bibfield  {title}
  {\bibinfo {title} {Solving quantum impurity problems in and out of
  equilibrium with the variational approach},\ }\href
  {https://doi.org/10.1103/PhysRevLett.121.026805} {\bibfield  {journal}
  {\bibinfo  {journal} {Phys. Rev. Lett.}\ }\textbf {\bibinfo {volume} {121}},\
  \bibinfo {pages} {026805} (\bibinfo {year} {2018})}\BibitemShut {NoStop}%
\bibitem [{\citenamefont {Shi}\ \emph {et~al.}(2018)\citenamefont {Shi},
  \citenamefont {Demler},\ and\ \citenamefont {{Ignacio Cirac}}}]{SHI2018245}%
  \BibitemOpen
  \bibfield  {author} {\bibinfo {author} {\bibfnamefont {T.}~\bibnamefont
  {Shi}}, \bibinfo {author} {\bibfnamefont {E.}~\bibnamefont {Demler}},\ and\
  \bibinfo {author} {\bibfnamefont {J.}~\bibnamefont {{Ignacio Cirac}}},\
  }\bibfield  {title} {\bibinfo {title} {Variational study of fermionic and
  bosonic systems with non-gaussian states: Theory and applications},\ }\href
  {https://doi.org/https://doi.org/10.1016/j.aop.2017.11.014} {\bibfield
  {journal} {\bibinfo  {journal} {Annals of Physics}\ }\textbf {\bibinfo
  {volume} {390}},\ \bibinfo {pages} {245} (\bibinfo {year}
  {2018})}\BibitemShut {NoStop}%
\bibitem [{\citenamefont {Cohen}\ \emph {et~al.}(2015)\citenamefont {Cohen},
  \citenamefont {Gull}, \citenamefont {Reichman},\ and\ \citenamefont
  {Millis}}]{cohen15taming}%
  \BibitemOpen
  \bibfield  {author} {\bibinfo {author} {\bibfnamefont {G.}~\bibnamefont
  {Cohen}}, \bibinfo {author} {\bibfnamefont {E.}~\bibnamefont {Gull}},
  \bibinfo {author} {\bibfnamefont {D.~R.}\ \bibnamefont {Reichman}},\ and\
  \bibinfo {author} {\bibfnamefont {A.~J.}\ \bibnamefont {Millis}},\ }\bibfield
   {title} {\bibinfo {title} {Taming the dynamical sign problem in real-time
  evolution of quantum many-body problems},\ }\href
  {https://doi.org/10.1103/PhysRevLett.115.266802} {\bibfield  {journal}
  {\bibinfo  {journal} {Phys. Rev. Lett.}\ }\textbf {\bibinfo {volume} {115}},\
  \bibinfo {pages} {266802} (\bibinfo {year} {2015})}\BibitemShut {NoStop}%
\bibitem [{\citenamefont {Bertrand}\ \emph {et~al.}(2019)\citenamefont
  {Bertrand}, \citenamefont {Florens}, \citenamefont {Parcollet},\ and\
  \citenamefont {Waintal}}]{Bertrand19Reconstructing}%
  \BibitemOpen
  \bibfield  {author} {\bibinfo {author} {\bibfnamefont {C.}~\bibnamefont
  {Bertrand}}, \bibinfo {author} {\bibfnamefont {S.}~\bibnamefont {Florens}},
  \bibinfo {author} {\bibfnamefont {O.}~\bibnamefont {Parcollet}},\ and\
  \bibinfo {author} {\bibfnamefont {X.}~\bibnamefont {Waintal}},\ }\bibfield
  {title} {\bibinfo {title} {Reconstructing nonequilibrium regimes of quantum
  many-body systems from the analytical structure of perturbative expansions},\
  }\href {https://doi.org/10.1103/PhysRevX.9.041008} {\bibfield  {journal}
  {\bibinfo  {journal} {Phys. Rev. X}\ }\textbf {\bibinfo {volume} {9}},\
  \bibinfo {pages} {041008} (\bibinfo {year} {2019})}\BibitemShut {NoStop}%
\bibitem [{\citenamefont {Ma\ifmmode~\check{c}\else \v{c}\fi{}ek}\ \emph
  {et~al.}(2020)\citenamefont {Ma\ifmmode~\check{c}\else \v{c}\fi{}ek},
  \citenamefont {Dumitrescu}, \citenamefont {Bertrand}, \citenamefont {Triggs},
  \citenamefont {Parcollet},\ and\ \citenamefont {Waintal}}]{QQMC20}%
  \BibitemOpen
  \bibfield  {author} {\bibinfo {author} {\bibfnamefont {M.}~\bibnamefont
  {Ma\ifmmode~\check{c}\else \v{c}\fi{}ek}}, \bibinfo {author} {\bibfnamefont
  {P.~T.}\ \bibnamefont {Dumitrescu}}, \bibinfo {author} {\bibfnamefont
  {C.}~\bibnamefont {Bertrand}}, \bibinfo {author} {\bibfnamefont
  {B.}~\bibnamefont {Triggs}}, \bibinfo {author} {\bibfnamefont
  {O.}~\bibnamefont {Parcollet}},\ and\ \bibinfo {author} {\bibfnamefont
  {X.}~\bibnamefont {Waintal}},\ }\bibfield  {title} {\bibinfo {title} {Quantum
  quasi-monte carlo technique for many-body perturbative expansions},\ }\href
  {https://doi.org/10.1103/PhysRevLett.125.047702} {\bibfield  {journal}
  {\bibinfo  {journal} {Phys. Rev. Lett.}\ }\textbf {\bibinfo {volume} {125}},\
  \bibinfo {pages} {047702} (\bibinfo {year} {2020})}\BibitemShut {NoStop}%
\bibitem [{\citenamefont {Ba\~nuls}\ \emph {et~al.}(2009)\citenamefont
  {Ba\~nuls}, \citenamefont {Hastings}, \citenamefont {Verstraete},\ and\
  \citenamefont {Cirac}}]{Banuls09}%
  \BibitemOpen
  \bibfield  {author} {\bibinfo {author} {\bibfnamefont {M.~C.}\ \bibnamefont
  {Ba\~nuls}}, \bibinfo {author} {\bibfnamefont {M.~B.}\ \bibnamefont
  {Hastings}}, \bibinfo {author} {\bibfnamefont {F.}~\bibnamefont
  {Verstraete}},\ and\ \bibinfo {author} {\bibfnamefont {J.~I.}\ \bibnamefont
  {Cirac}},\ }\bibfield  {title} {\bibinfo {title} {Matrix product states for
  dynamical simulation of infinite chains},\ }\href
  {https://doi.org/10.1103/PhysRevLett.102.240603} {\bibfield  {journal}
  {\bibinfo  {journal} {Phys. Rev. Lett.}\ }\textbf {\bibinfo {volume} {102}},\
  \bibinfo {pages} {240603} (\bibinfo {year} {2009})}\BibitemShut {NoStop}%
\bibitem [{\citenamefont {Huang}\ \emph {et~al.}(2014)\citenamefont {Huang},
  \citenamefont {Chen}, \citenamefont {Kao},\ and\ \citenamefont
  {Xiang}}]{huang14longtime}%
  \BibitemOpen
  \bibfield  {author} {\bibinfo {author} {\bibfnamefont {Y.-K.}\ \bibnamefont
  {Huang}}, \bibinfo {author} {\bibfnamefont {P.}~\bibnamefont {Chen}},
  \bibinfo {author} {\bibfnamefont {Y.-J.}\ \bibnamefont {Kao}},\ and\ \bibinfo
  {author} {\bibfnamefont {T.}~\bibnamefont {Xiang}},\ }\bibfield  {title}
  {\bibinfo {title} {Long-time dynamics of quantum chains: Transfer-matrix
  renormalization group and entanglement of the maximal eigenvector},\ }\href
  {https://doi.org/10.1103/PhysRevB.89.201102} {\bibfield  {journal} {\bibinfo
  {journal} {Phys. Rev. B}\ }\textbf {\bibinfo {volume} {89}},\ \bibinfo
  {pages} {201102} (\bibinfo {year} {2014})}\BibitemShut {NoStop}%
\bibitem [{\citenamefont {Lerose}\ \emph
  {et~al.}(2021{\natexlab{a}})\citenamefont {Lerose}, \citenamefont {Sonner},\
  and\ \citenamefont {Abanin}}]{lerose2020}%
  \BibitemOpen
  \bibfield  {author} {\bibinfo {author} {\bibfnamefont {A.}~\bibnamefont
  {Lerose}}, \bibinfo {author} {\bibfnamefont {M.}~\bibnamefont {Sonner}},\
  and\ \bibinfo {author} {\bibfnamefont {D.~A.}\ \bibnamefont {Abanin}},\
  }\bibfield  {title} {\bibinfo {title} {Influence matrix approach to many-body
  floquet dynamics},\ }\href {https://doi.org/10.1103/PhysRevX.11.021040}
  {\bibfield  {journal} {\bibinfo  {journal} {Phys. Rev. X}\ }\textbf {\bibinfo
  {volume} {11}},\ \bibinfo {pages} {021040} (\bibinfo {year}
  {2021}{\natexlab{a}})}\BibitemShut {NoStop}%
\bibitem [{\citenamefont {Lerose}\ \emph
  {et~al.}(2021{\natexlab{b}})\citenamefont {Lerose}, \citenamefont {Sonner},\
  and\ \citenamefont {Abanin}}]{lerose2021}%
  \BibitemOpen
  \bibfield  {author} {\bibinfo {author} {\bibfnamefont {A.}~\bibnamefont
  {Lerose}}, \bibinfo {author} {\bibfnamefont {M.}~\bibnamefont {Sonner}},\
  and\ \bibinfo {author} {\bibfnamefont {D.~A.}\ \bibnamefont {Abanin}},\
  }\bibfield  {title} {\bibinfo {title} {Scaling of temporal entanglement in
  proximity to integrability},\ }\href
  {https://doi.org/10.1103/PhysRevB.104.035137} {\bibfield  {journal} {\bibinfo
   {journal} {Phys. Rev. B}\ }\textbf {\bibinfo {volume} {104}},\ \bibinfo
  {pages} {035137} (\bibinfo {year} {2021}{\natexlab{b}})}\BibitemShut
  {NoStop}%
\bibitem [{\citenamefont {Ye}\ and\ \citenamefont {Chan}(2021)}]{Chan21}%
  \BibitemOpen
  \bibfield  {author} {\bibinfo {author} {\bibfnamefont {E.}~\bibnamefont
  {Ye}}\ and\ \bibinfo {author} {\bibfnamefont {G.~K.-L.}\ \bibnamefont
  {Chan}},\ }\bibfield  {title} {\bibinfo {title} {Constructing tensor network
  influence functionals for general quantum dynamics},\ }\href
  {https://doi.org/10.1063/5.0047260} {\bibfield  {journal} {\bibinfo
  {journal} {The Journal of Chemical Physics}\ }\textbf {\bibinfo {volume}
  {155}},\ \bibinfo {pages} {044104} (\bibinfo {year} {2021})},\ \Eprint
  {https://arxiv.org/abs/https://doi.org/10.1063/5.0047260}
  {https://doi.org/10.1063/5.0047260} \BibitemShut {NoStop}%
\bibitem [{\citenamefont {Sonner}\ \emph {et~al.}(2021)\citenamefont {Sonner},
  \citenamefont {Lerose},\ and\ \citenamefont {Abanin}}]{sonner2021influence}%
  \BibitemOpen
  \bibfield  {author} {\bibinfo {author} {\bibfnamefont {M.}~\bibnamefont
  {Sonner}}, \bibinfo {author} {\bibfnamefont {A.}~\bibnamefont {Lerose}},\
  and\ \bibinfo {author} {\bibfnamefont {D.~A.}\ \bibnamefont {Abanin}},\
  }\bibfield  {title} {\bibinfo {title} {Influence functional of many-body
  systems: Temporal entanglement and matrix-product state representation},\
  }\href@noop {} {\bibfield  {journal} {\bibinfo  {journal} {Annals of
  Physics}\ }\textbf {\bibinfo {volume} {435}},\ \bibinfo {pages} {168677}
  (\bibinfo {year} {2021})}\BibitemShut {NoStop}%
\bibitem [{\citenamefont {Sonner}\ \emph {et~al.}(2022)\citenamefont {Sonner},
  \citenamefont {Lerose},\ and\ \citenamefont
  {Abanin}}]{sonner22characterizing}%
  \BibitemOpen
  \bibfield  {author} {\bibinfo {author} {\bibfnamefont {M.}~\bibnamefont
  {Sonner}}, \bibinfo {author} {\bibfnamefont {A.}~\bibnamefont {Lerose}},\
  and\ \bibinfo {author} {\bibfnamefont {D.~A.}\ \bibnamefont {Abanin}},\
  }\bibfield  {title} {\bibinfo {title} {Characterizing many-body localization
  via exact disorder-averaged quantum noise},\ }\href
  {https://doi.org/10.1103/PhysRevB.105.L020203} {\bibfield  {journal}
  {\bibinfo  {journal} {Phys. Rev. B}\ }\textbf {\bibinfo {volume} {105}},\
  \bibinfo {pages} {L020203} (\bibinfo {year} {2022})}\BibitemShut {NoStop}%
\bibitem [{\citenamefont {Piroli}\ \emph {et~al.}(2020)\citenamefont {Piroli},
  \citenamefont {Bertini}, \citenamefont {Cirac},\ and\ \citenamefont
  {Prosen}}]{Piroli2020}%
  \BibitemOpen
  \bibfield  {author} {\bibinfo {author} {\bibfnamefont {L.}~\bibnamefont
  {Piroli}}, \bibinfo {author} {\bibfnamefont {B.}~\bibnamefont {Bertini}},
  \bibinfo {author} {\bibfnamefont {J.~I.}\ \bibnamefont {Cirac}},\ and\
  \bibinfo {author} {\bibfnamefont {T.~c.~v.}\ \bibnamefont {Prosen}},\
  }\bibfield  {title} {\bibinfo {title} {Exact dynamics in dual-unitary quantum
  circuits},\ }\href {https://doi.org/10.1103/PhysRevB.101.094304} {\bibfield
  {journal} {\bibinfo  {journal} {Phys. Rev. B}\ }\textbf {\bibinfo {volume}
  {101}},\ \bibinfo {pages} {094304} (\bibinfo {year} {2020})}\BibitemShut
  {NoStop}%
\bibitem [{\citenamefont {Klobas}\ \emph {et~al.}(2021)\citenamefont {Klobas},
  \citenamefont {Bertini},\ and\ \citenamefont {Piroli}}]{klobas2021exact}%
  \BibitemOpen
  \bibfield  {author} {\bibinfo {author} {\bibfnamefont {K.}~\bibnamefont
  {Klobas}}, \bibinfo {author} {\bibfnamefont {B.}~\bibnamefont {Bertini}},\
  and\ \bibinfo {author} {\bibfnamefont {L.}~\bibnamefont {Piroli}},\
  }\bibfield  {title} {\bibinfo {title} {Exact thermalization dynamics in the
  “rule 54” quantum cellular automaton},\ }\href@noop {} {\bibfield
  {journal} {\bibinfo  {journal} {Physical Review Letters}\ }\textbf {\bibinfo
  {volume} {126}},\ \bibinfo {pages} {160602} (\bibinfo {year}
  {2021})}\BibitemShut {NoStop}%
\bibitem [{\citenamefont {Giudice}\ \emph {et~al.}(2022)\citenamefont
  {Giudice}, \citenamefont {Giudici}, \citenamefont {Sonner}, \citenamefont
  {Thoenniss}, \citenamefont {Lerose}, \citenamefont {Abanin},\ and\
  \citenamefont {Piroli}}]{giudice2021temporal}%
  \BibitemOpen
  \bibfield  {author} {\bibinfo {author} {\bibfnamefont {G.}~\bibnamefont
  {Giudice}}, \bibinfo {author} {\bibfnamefont {G.}~\bibnamefont {Giudici}},
  \bibinfo {author} {\bibfnamefont {M.}~\bibnamefont {Sonner}}, \bibinfo
  {author} {\bibfnamefont {J.}~\bibnamefont {Thoenniss}}, \bibinfo {author}
  {\bibfnamefont {A.}~\bibnamefont {Lerose}}, \bibinfo {author} {\bibfnamefont
  {D.~A.}\ \bibnamefont {Abanin}},\ and\ \bibinfo {author} {\bibfnamefont
  {L.}~\bibnamefont {Piroli}},\ }\bibfield  {title} {\bibinfo {title} {Temporal
  entanglement, quasiparticles, and the role of interactions},\ }\href
  {https://doi.org/10.1103/PhysRevLett.128.220401} {\bibfield  {journal}
  {\bibinfo  {journal} {Phys. Rev. Lett.}\ }\textbf {\bibinfo {volume} {128}},\
  \bibinfo {pages} {220401} (\bibinfo {year} {2022})}\BibitemShut {NoStop}%
\bibitem [{\citenamefont {Lerose}\ \emph {et~al.}(2022)\citenamefont {Lerose},
  \citenamefont {Sonner},\ and\ \citenamefont {Abanin}}]{lerose2022overcoming}%
  \BibitemOpen
  \bibfield  {author} {\bibinfo {author} {\bibfnamefont {A.}~\bibnamefont
  {Lerose}}, \bibinfo {author} {\bibfnamefont {M.}~\bibnamefont {Sonner}},\
  and\ \bibinfo {author} {\bibfnamefont {D.~A.}\ \bibnamefont {Abanin}},\
  }\bibfield  {title} {\bibinfo {title} {Overcoming the entanglement barrier in
  quantum many-body dynamics via space-time duality},\ }\href@noop {}
  {\bibfield  {journal} {\bibinfo  {journal} {arXiv preprint arXiv:2201.04150}\
  } (\bibinfo {year} {2022})}\BibitemShut {NoStop}%
\bibitem [{\citenamefont {{Strathearn}}\ \emph {et~al.}(2018)\citenamefont
  {{Strathearn}}, \citenamefont {{Kirton}}, \citenamefont {{Kilda}},
  \citenamefont {{Keeling}},\ and\ \citenamefont {{Lovett}}}]{TEMPO}%
  \BibitemOpen
  \bibfield  {author} {\bibinfo {author} {\bibfnamefont {A.}~\bibnamefont
  {{Strathearn}}}, \bibinfo {author} {\bibfnamefont {P.}~\bibnamefont
  {{Kirton}}}, \bibinfo {author} {\bibfnamefont {D.}~\bibnamefont {{Kilda}}},
  \bibinfo {author} {\bibfnamefont {J.}~\bibnamefont {{Keeling}}},\ and\
  \bibinfo {author} {\bibfnamefont {B.~W.}\ \bibnamefont {{Lovett}}},\
  }\bibfield  {title} {\bibinfo {title} {{Efficient non-Markovian quantum
  dynamics using time-evolving matrix product operators}},\ }\href
  {https://doi.org/10.1038/s41467-018-05617-3} {\bibfield  {journal} {\bibinfo
  {journal} {Nature Communications}\ }\textbf {\bibinfo {volume} {9}},\
  \bibinfo {eid} {3322} (\bibinfo {year} {2018})}\BibitemShut {NoStop}%
\bibitem [{\citenamefont {Thoenniss}\ \emph {et~al.}(2022)\citenamefont
  {Thoenniss}, \citenamefont {Lerose},\ and\ \citenamefont
  {Abanin}}]{Thoenniss22}%
  \BibitemOpen
  \bibfield  {author} {\bibinfo {author} {\bibfnamefont {J.}~\bibnamefont
  {Thoenniss}}, \bibinfo {author} {\bibfnamefont {A.}~\bibnamefont {Lerose}},\
  and\ \bibinfo {author} {\bibfnamefont {D.~A.}\ \bibnamefont {Abanin}},\
  }\bibfield  {title} {\bibinfo {title} {Non-equilibrium quantum impurity
  problems via matrix-product states in the temporal domain}\ }\href
  {https://doi.org/10.48550/ARXIV.2205.04995} {10.48550/ARXIV.2205.04995}
  (\bibinfo {year} {2022})\BibitemShut {NoStop}%
\bibitem [{\citenamefont {Bose}\ and\ \citenamefont
  {Walters}(2021)}]{bose2021tensor}%
  \BibitemOpen
  \bibfield  {author} {\bibinfo {author} {\bibfnamefont {A.}~\bibnamefont
  {Bose}}\ and\ \bibinfo {author} {\bibfnamefont {P.~L.}\ \bibnamefont
  {Walters}},\ }\bibfield  {title} {\bibinfo {title} {A tensor network
  representation of path integrals: Implementation and analysis},\ }\href@noop
  {} {\bibfield  {journal} {\bibinfo  {journal} {arXiv preprint
  arXiv:2106.12523}\ } (\bibinfo {year} {2021})}\BibitemShut {NoStop}%
\bibitem [{\citenamefont {Feynman}\ and\ \citenamefont
  {Vernon}(1963)}]{FeynmanVernon}%
  \BibitemOpen
  \bibfield  {author} {\bibinfo {author} {\bibfnamefont {R.}~\bibnamefont
  {Feynman}}\ and\ \bibinfo {author} {\bibfnamefont {F.}~\bibnamefont
  {Vernon}},\ }\bibfield  {title} {\bibinfo {title} {The theory of a general
  quantum system interacting with a linear dissipative system},\ }\href
  {https://doi.org/https://doi.org/10.1016/0003-4916(63)90068-X} {\bibfield
  {journal} {\bibinfo  {journal} {Annals of Physics}\ }\textbf {\bibinfo
  {volume} {24}},\ \bibinfo {pages} {118 } (\bibinfo {year}
  {1963})}\BibitemShut {NoStop}%
\bibitem [{\citenamefont {Bravyi}\ and\ \citenamefont
  {Gosset}(2017)}]{bravyi2017complexity}%
  \BibitemOpen
  \bibfield  {author} {\bibinfo {author} {\bibfnamefont {S.}~\bibnamefont
  {Bravyi}}\ and\ \bibinfo {author} {\bibfnamefont {D.}~\bibnamefont
  {Gosset}},\ }\bibfield  {title} {\bibinfo {title} {Complexity of quantum
  impurity problems},\ }\href@noop {} {\bibfield  {journal} {\bibinfo
  {journal} {Communications in Mathematical Physics}\ }\textbf {\bibinfo
  {volume} {356}},\ \bibinfo {pages} {451} (\bibinfo {year}
  {2017})}\BibitemShut {NoStop}%
\bibitem [{\citenamefont {Debertolis}\ \emph {et~al.}(2021)\citenamefont
  {Debertolis}, \citenamefont {Florens},\ and\ \citenamefont
  {Snyman}}]{Debertolis21fewbody}%
  \BibitemOpen
  \bibfield  {author} {\bibinfo {author} {\bibfnamefont {M.}~\bibnamefont
  {Debertolis}}, \bibinfo {author} {\bibfnamefont {S.}~\bibnamefont
  {Florens}},\ and\ \bibinfo {author} {\bibfnamefont {I.}~\bibnamefont
  {Snyman}},\ }\bibfield  {title} {\bibinfo {title} {Few-body nature of kondo
  correlated ground states},\ }\href
  {https://doi.org/10.1103/PhysRevB.103.235166} {\bibfield  {journal} {\bibinfo
   {journal} {Phys. Rev. B}\ }\textbf {\bibinfo {volume} {103}},\ \bibinfo
  {pages} {235166} (\bibinfo {year} {2021})}\BibitemShut {NoStop}%
\bibitem [{\citenamefont {Fishman}\ and\ \citenamefont
  {White}(2015)}]{Fishman2015MPS}%
  \BibitemOpen
  \bibfield  {author} {\bibinfo {author} {\bibfnamefont {M.~T.}\ \bibnamefont
  {Fishman}}\ and\ \bibinfo {author} {\bibfnamefont {S.~R.}\ \bibnamefont
  {White}},\ }\bibfield  {title} {\bibinfo {title} {Compression of correlation
  matrices and an efficient method for forming matrix product states of
  fermionic gaussian states},\ }\href
  {https://doi.org/10.1103/PhysRevB.92.075132} {\bibfield  {journal} {\bibinfo
  {journal} {Phys. Rev. B}\ }\textbf {\bibinfo {volume} {92}},\ \bibinfo
  {pages} {075132} (\bibinfo {year} {2015})}\BibitemShut {NoStop}%
\bibitem [{\citenamefont {Schuch}\ and\ \citenamefont
  {Bauer}(2019)}]{SchuchGaussianMPS}%
  \BibitemOpen
  \bibfield  {author} {\bibinfo {author} {\bibfnamefont {N.}~\bibnamefont
  {Schuch}}\ and\ \bibinfo {author} {\bibfnamefont {B.}~\bibnamefont {Bauer}},\
  }\bibfield  {title} {\bibinfo {title} {Matrix product state algorithms for
  gaussian fermionic states},\ }\href
  {https://doi.org/10.1103/PhysRevB.100.245121} {\bibfield  {journal} {\bibinfo
   {journal} {Phys. Rev. B}\ }\textbf {\bibinfo {volume} {100}},\ \bibinfo
  {pages} {245121} (\bibinfo {year} {2019})}\BibitemShut {NoStop}%
\bibitem [{\citenamefont {Medvedyeva}\ \emph {et~al.}(2013)\citenamefont
  {Medvedyeva}, \citenamefont {Hoffmann},\ and\ \citenamefont
  {Kehrein}}]{Medvedyeva13SpatiotemporalBuildup}%
  \BibitemOpen
  \bibfield  {author} {\bibinfo {author} {\bibfnamefont {M.}~\bibnamefont
  {Medvedyeva}}, \bibinfo {author} {\bibfnamefont {A.}~\bibnamefont
  {Hoffmann}},\ and\ \bibinfo {author} {\bibfnamefont {S.}~\bibnamefont
  {Kehrein}},\ }\bibfield  {title} {\bibinfo {title} {Spatiotemporal buildup of
  the kondo screening cloud},\ }\href
  {https://doi.org/10.1103/PhysRevB.88.094306} {\bibfield  {journal} {\bibinfo
  {journal} {Phys. Rev. B}\ }\textbf {\bibinfo {volume} {88}},\ \bibinfo
  {pages} {094306} (\bibinfo {year} {2013})}\BibitemShut {NoStop}%
\bibitem [{\citenamefont {Schmidt}\ \emph {et~al.}(2008)\citenamefont
  {Schmidt}, \citenamefont {Werner}, \citenamefont {M\"uhlbacher},\ and\
  \citenamefont {Komnik}}]{schmidt08transient}%
  \BibitemOpen
  \bibfield  {author} {\bibinfo {author} {\bibfnamefont {T.~L.}\ \bibnamefont
  {Schmidt}}, \bibinfo {author} {\bibfnamefont {P.}~\bibnamefont {Werner}},
  \bibinfo {author} {\bibfnamefont {L.}~\bibnamefont {M\"uhlbacher}},\ and\
  \bibinfo {author} {\bibfnamefont {A.}~\bibnamefont {Komnik}},\ }\bibfield
  {title} {\bibinfo {title} {Transient dynamics of the anderson impurity model
  out of equilibrium},\ }\href {https://doi.org/10.1103/PhysRevB.78.235110}
  {\bibfield  {journal} {\bibinfo  {journal} {Phys. Rev. B}\ }\textbf {\bibinfo
  {volume} {78}},\ \bibinfo {pages} {235110} (\bibinfo {year}
  {2008})}\BibitemShut {NoStop}%
\bibitem [{\citenamefont {Werner}\ \emph {et~al.}(2010)\citenamefont {Werner},
  \citenamefont {Oka}, \citenamefont {Eckstein},\ and\ \citenamefont
  {Millis}}]{werner10weakcoupling}%
  \BibitemOpen
  \bibfield  {author} {\bibinfo {author} {\bibfnamefont {P.}~\bibnamefont
  {Werner}}, \bibinfo {author} {\bibfnamefont {T.}~\bibnamefont {Oka}},
  \bibinfo {author} {\bibfnamefont {M.}~\bibnamefont {Eckstein}},\ and\
  \bibinfo {author} {\bibfnamefont {A.~J.}\ \bibnamefont {Millis}},\ }\bibfield
   {title} {\bibinfo {title} {Weak-coupling quantum monte carlo calculations on
  the keldysh contour: Theory and application to the current-voltage
  characteristics of the anderson model},\ }\href
  {https://doi.org/10.1103/PhysRevB.81.035108} {\bibfield  {journal} {\bibinfo
  {journal} {Phys. Rev. B}\ }\textbf {\bibinfo {volume} {81}},\ \bibinfo
  {pages} {035108} (\bibinfo {year} {2010})}\BibitemShut {NoStop}%
\bibitem [{\citenamefont {Fugger}\ \emph {et~al.}(2018)\citenamefont {Fugger},
  \citenamefont {Dorda}, \citenamefont {Schwarz}, \citenamefont {von Delft},\
  and\ \citenamefont {Arrigoni}}]{Fugger_2018}%
  \BibitemOpen
  \bibfield  {author} {\bibinfo {author} {\bibfnamefont {D.~M.}\ \bibnamefont
  {Fugger}}, \bibinfo {author} {\bibfnamefont {A.}~\bibnamefont {Dorda}},
  \bibinfo {author} {\bibfnamefont {F.}~\bibnamefont {Schwarz}}, \bibinfo
  {author} {\bibfnamefont {J.}~\bibnamefont {von Delft}},\ and\ \bibinfo
  {author} {\bibfnamefont {E.}~\bibnamefont {Arrigoni}},\ }\bibfield  {title}
  {\bibinfo {title} {Nonequilibrium kondo effect in a magnetic field: auxiliary
  master equation approach},\ }\href {https://doi.org/10.1088/1367-2630/aa9fdc}
  {\bibfield  {journal} {\bibinfo  {journal} {New Journal of Physics}\ }\textbf
  {\bibinfo {volume} {20}},\ \bibinfo {pages} {013030} (\bibinfo {year}
  {2018})}\BibitemShut {NoStop}%
\bibitem [{Note1()}]{Note1}%
  \BibitemOpen
  \bibinfo {note} {The choice of an environment defined by unitary evolution
  allows us to avoid errors associated with time-discretization of a
  pre-defined spectral density $\Gamma (\omega )$.}\BibitemShut {Stop}%
\bibitem [{\citenamefont {Ng}\ \emph {et~al.}(2022)\citenamefont {Ng},
  \citenamefont {Park}, \citenamefont {Millis}, \citenamefont {Chan},\ and\
  \citenamefont {Reichman}}]{Ng22Realtime}%
  \BibitemOpen
  \bibfield  {author} {\bibinfo {author} {\bibfnamefont {N.}~\bibnamefont
  {Ng}}, \bibinfo {author} {\bibfnamefont {G.}~\bibnamefont {Park}}, \bibinfo
  {author} {\bibfnamefont {A.~J.}\ \bibnamefont {Millis}}, \bibinfo {author}
  {\bibfnamefont {G.~K.-L.}\ \bibnamefont {Chan}},\ and\ \bibinfo {author}
  {\bibfnamefont {D.~R.}\ \bibnamefont {Reichman}},\ }\href
  {https://doi.org/10.48550/ARXIV.2211.10430} {\bibinfo {title} {Real time
  evolution of anderson impurity models via tensor network influence
  functionals}} (\bibinfo {year} {2022})\BibitemShut {NoStop}%
\bibitem [{\citenamefont {Wilde}(2013)}]{wilde2013quantum}%
  \BibitemOpen
  \bibfield  {author} {\bibinfo {author} {\bibfnamefont {M.~M.}\ \bibnamefont
  {Wilde}},\ }\href@noop {} {\emph {\bibinfo {title} {Quantum information
  theory}}}\ (\bibinfo  {publisher} {Cambridge University Press},\ \bibinfo
  {year} {2013})\BibitemShut {NoStop}%
\bibitem [{foo()}]{footnote_sm_superoperator}%
  \BibitemOpen
  \href@noop {} {\bibinfo  {journal} {In the strict sense,
  $\{A_\sigma(m)\}_{m=1}^M$ are not superoperators as their bond dimension (and
  thus the equivalent of the environment operator space) is not guaranteed to
  be a square integer, and CPTP property is not enforced. Furthermore, this
  property would require to fix the gauge freedom of the MPS}\ }\BibitemShut
  {NoStop}%
\end{thebibliography}%
\end{document}


\title{Supplemental Material: \\ An efficient method for quantum impurity problems out of equilibrium}

\author{Julian Thoenniss}
\affiliation{Department of Theoretical Physics,
University of Geneva, Quai Ernest-Ansermet 30,
1205 Geneva, Switzerland}

\author{Michael  Sonner}
\affiliation{Department of Theoretical Physics,
University of Geneva, Quai Ernest-Ansermet 30,
1205 Geneva, Switzerland}

\author{Alessio Lerose}
\affiliation{Department of Theoretical Physics,
University of Geneva, Quai Ernest-Ansermet 30,
1205 Geneva, Switzerland}

\author{Dmitry A. Abanin}
\affiliation{Department of Theoretical Physics,
University of Geneva, Quai Ernest-Ansermet 30,
1205 Geneva, Switzerland}

\date{\today}

\maketitle
\appendix
\section{Derivation of Eq.~(4)
}
\label{app:overlap}
We start by recalling the standard derivation of the path integral in Eq.~(2
).
Defining the evolution operator $U=\exp(i\,\delta t\, H)$ for a time step $\delta t = T/M$ ($M\gg 1$) and the Hamiltonian $H$ from Eq.~(1
), the expectation value of an impurity observable can be expressed as
\begin{multline}
\langle \hat{O}(t_{m^*}) \rangle \\
= \text{Tr}_\text{imp}\Bigg[   \text{Tr}_\text{bath}\Big(\, U^{M-m^*} \, \hat{O} \, U^{m^*}  \big( \rho_\text{imp} \otimes \rho_\text{bath} \big)
 (U^\dagger)^M \, \Big) \Bigg].
 \label{eq:stextbook_trotter}
 \end{multline}  Here, $t_{m^*}= m^* \cdot \delta t$ denotes a point on the discrete-time lattice and $m^* \in \{0,1,\hdots, M\}$.
This expression is cast into path integral form by inserting Grassmann resolutions of identity $\mathds{1}_\tau=\otimes_\sigma \mathds{1}_{\sigma,\tau},$ where
\begin{multline}
    \mathds{1}_{\sigma,\tau} =  \int d(\bar{\eta}_{\sigma,\tau},\eta_{\sigma,\tau})d(\bm{\bar{\xi}}_{\sigma,\tau},\bm{\xi}_{\sigma,\tau})\\ e^{-\bar{\eta}_{\sigma,\tau} \eta_{\sigma,\tau}- \bar{\bm{\xi}}_{\sigma,\tau}\bm{\xi}_{\sigma,\tau}}| \eta_{\sigma,\tau}, \bm{\xi}_{\sigma,\tau}\rangle \langle \bar{\eta}_{\sigma,\tau},\bar{\bm{\xi}}_{\sigma,\tau}|,
    \label{eq:GM_identity}
\end{multline} between every multiplication of operators. Here, $\bar{\eta}_{\sigma,\tau},\eta_{\sigma,\tau}$ are impurity variables with index $\tau=0^\pm,\dots,M^\pm$ on the (discretized) Keldysh contour, while $\bar{\bm{\xi}}_{\sigma,\tau}=(\bar{\xi}_{j=1,\sigma,\tau},\hdots,\bar{\xi}_{j=L,\sigma,\tau})^T,\,\bm{\xi}_{\sigma,\tau} = (\xi_{j=1,\sigma,\tau},\hdots,\xi_{j=L,\sigma,\tau})^T$ are the degrees of freedom of the environment  (made of $L$ fermionic modes). In total, we thus insert $2(M+1)$ identity resolutions per fermionic mode. In the limit $\delta t \to 0,$ Eq.~(2
) is retrieved following standard textbook passages.

To define a clean prescription for our temporal wave functions overlap, however, it is more convenient to further split the evolution operator $U$ into a local impurity and environment+tunneling evolution operators, i.e. $U\approx  U_\text{imp}\cdot U_\text{hyb}$ (with the same error $\mathcal{O}(\delta t^2)$ as before). Here, we defined $U_\text{hyb}=\exp\big(i\,\delta t\, H_\text{hyb}\big)$ with 
\begin{equation}
\begin{split}
    H_\text{hyb} &= H - H_\text{imp} = \\ &= \sum_{\substack {k\\\sigma=\uparrow,\downarrow\\\alpha=L,R}} \Big[ \big( t_k d_\sigma^\dagger c_{k,\alpha,\sigma} + h.c.\big) + \epsilon_{k} c_{k,\alpha, \sigma}^\dagger c_{k,\alpha,\sigma}\Big],
\end{split}
\end{equation}
and, for later convenience, we choose
 \begin{equation}\label{eq:U_imp}
   U_\text{imp}= \begin{cases} e^{i\,\delta t H_\text{imp}}  & \text{ for evolution up to time } t_{m^*} \\
      \mathds{1}_\text{imp}& \text{ for evolution in range } [t_{m^*}, T] .
     \end{cases}
     \end{equation}
[While this choice is convenient for the conceptual derivation within the Grassmann formalism, we found that a slightly modified prescription, where we replace $\mathds{1}_\text{imp}$ in Eq.~(\ref{eq:U_imp}) by the perfectly depolarizing channel, is numerically more efficient, see App.~\ref{app:mps}.] In total, with this modified Trotter decomposition, we insert $4M$ Grassmann identity resolutions per spatial site and $\tau = 0^\pm,\dots,(2M-1)^\pm.$
In Fig.~\ref{fig:Grassmann_labels}, we illustrate how the resulting $8M$ Grassmann variables are associated with the legs of the IF and impurity tensors, respectively.

\setcounter{figure}{0}
\renewcommand{\figurename}{Figure}
\renewcommand{\thefigure}{S\arabic{figure}}
  \begin{figure}[t]
\includegraphics[width=0.34\textwidth]{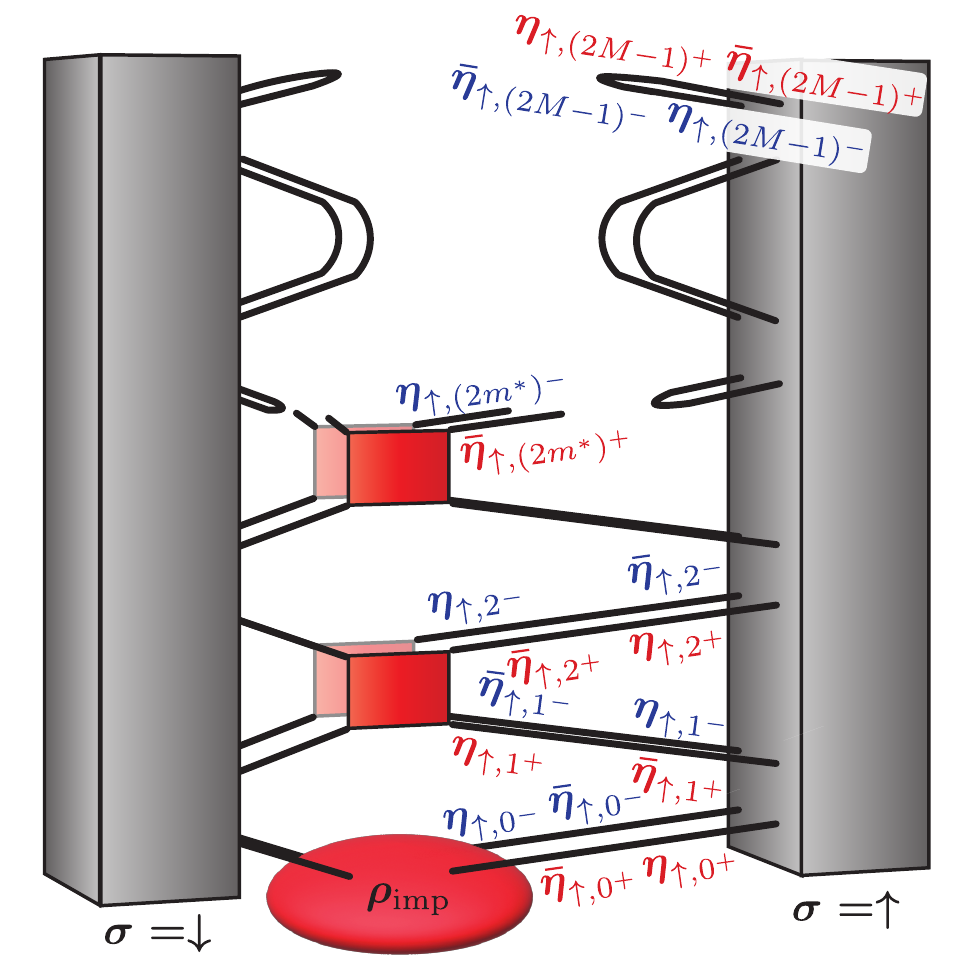}
\centering
\caption{After inserting Grassmann resolutions of identity into
    Eq.~(\ref{eq:stextbook_trotter}), each leg of the IF and impurity tensor
    gets associated with a Grassmann variable [red (blue) color refers to the
    forward (backward) Keldysh branch]. Here, we show the density matrix of the impurity after two time steps.
    It is obtained from an IM containing $M=4$ time steps. The impurity gates at time
    steps later than $m^*$ are replaced by the identity operator.}
\label{fig:Grassmann_labels}
\end{figure}
 
  In order to arrive at the overlap form of Eq.~(4
  ), we manipulate the path integral in a way that results in the following structure: All variables associated with the kernel of the spin-up (down) IF should be conjugate (non-conjugate) and opposite for the impurity kernel.
  This is achieved by making appropriate variable substitutions in the system-variables of the identity resolution, Eq.~(\ref{eq:GM_identity}). We define these modified identity resolutions as
  \begin{align}\nonumber
      \mathds{1}^\prime_{\sigma,\tau} \quad &\text{ with substitution } \bar{\eta}_{\sigma,\tau}\to \eta_{\sigma,\tau},\, \eta_{\sigma,\tau} \to -\bar{\eta}_{\sigma,\tau},\\
      \nonumber
   \mathds{1}^{\prime\prime}_{\sigma,\tau}\quad &\text{ with substitution } \bar{\eta}_{\sigma,\tau}\to -\eta_{\sigma,\tau},\, \eta_{\sigma,\tau} \to \bar{\eta}_{\sigma,\tau}.
  \end{align}
  With this, Grassmann identities are inserted between the hybridization- and impurity evolution operators on the forward branch in the following way: 
  \begin{equation}
  U_\text{imp}\cdot\mathds{1}_{(2m+1)^+} \cdot U_\text{hyb}\cdot\mathds{1}_{(2m)^+},\end{equation}
  with
  \begin{align}
     \mathds{1}_{(2m+1)^+} =& \mathds{1}_{\uparrow,(2m+1)^+}\otimes \mathds{1}^\prime_{\downarrow,(2m+1)^+},\\
    \mathds{1}_{(2m)^+}=&  \mathds{1}^{\prime\prime}_{\uparrow,(2m)^+}\otimes \mathds{1}_{\downarrow,(2m)^+}.
  \end{align} On the backward branch, we insert identities as follows:
    \begin{equation}
 \mathds{1}_{(2m)^-} \cdot U^\dagger_\text{hyb}\cdot \mathds{1}_{(2m+1)^-} U^\dagger_\text{imp},\end{equation} with
 \begin{align}
     \mathds{1}_{(2m)^-} =& \mathds{1}_{\uparrow,(2m)^-}\otimes \mathds{1}^\prime_{\downarrow,(2m)^-},\\
     \mathds{1}_{(2m+1)^-} =& \mathds{1}^{\prime\prime}_{\uparrow,(2m+1)^-}\otimes \mathds{1}_{\downarrow,(2m+1)^-}.
  \end{align} 
  With these insertions, one arrives at Eq.~(4
  ). Note that these variable substitutions alter the signs of some components of the impurity kernel, while they amount to a simple renaming of variables for the IF. 
  
  The resulting discrete-time IF has Gaussian form,
  \begin{equation}
   \mathcal{I}[\bm{\eta}_{\sigma}] =\exp\Big(\sum_{m,m^\prime}
\bm{\eta}^T_{\sigma,m} \mathbf{B}_{mm^\prime}
\bm{\eta}_{\sigma,m^\prime}\Big),
\label{eq:IF_form}
\end{equation} with
$$\bm{\eta}_{\sigma,m }= (\eta_{\sigma,(2m)^+} ,\eta_{\sigma,(2m)^-} ,\eta_{\sigma,(2m+1)^+} ,\eta_{\sigma,(2m+1)^-})^T.$$
 The matrix $\mathbf{B}$ that appears here is the exact Gaussian influence action of the trotterized (Floquet) environment~[52].
  To understand its relation to the continuous-time result, it is convenient to express it in terms of a discrete-time hybridization matrix $\mathbf{\Delta}$,
\begin{multline}
\mathcal{I}[\bm{\eta}_\sigma]=\exp\Bigg[ \sum_{m\geq m^\prime} 
\bm{\eta}_{\sigma,m}^T \mathbf{\Delta}_{mm^\prime}
\bm{\eta}_{\sigma,m}\Bigg]\, e^{\eta_{\sigma,0^+} \eta_{\sigma,0^-}}\\
\times e^{ \sum_{m=0}^{M-1} \big( \eta_{\sigma,(2m+1)^+} \eta_{\sigma,(2m)^+} + \eta_{\sigma,(2m)^-} \eta_{\sigma,(2m+1)^-} \big)},
\label{eq:IF_overlap_terms}
\end{multline}
where the terms in the second and third exponential in Eq.~(\ref{eq:IF_overlap_terms}) stem from the overlap of Grassmann coherent states. 

For a general discrete-time (Floquet) unitary evolution of the environment, the matrix $\mathbf{\Delta}$ has a complicated structure, which we evaluated exactly e.g. for setups with chain-environments [52]. 
However, the structure greatly simplifies in the Trotter limit $\delta t\to 0,$
where  $\mathbf{\Delta}$ can be written in the form
\begin{equation}
    \mathbf{\Delta}_{mm^\prime} =(\delta t)^2 \bigg[\sum_\alpha \int \frac{d\omega}{2\pi}\,\Gamma(\omega)\, \mathbf{G}^\alpha_{mm^\prime}(\omega) \;+ \mathcal{O}\big(\delta t\big) \bigg].
    \label{eq_discreteDelta}
\end{equation} 
Here, $\mathbf{G}_{mm^\prime}^\alpha(\omega)$ is a matrix of non-interacting  Green's functions of the environment. 
  For fermion-number-conserving Hamiltonians as considered in this work, this yields (omitting $\omega$-dependence for simplicity)
\begin{align}
    \mathbf{G}^\alpha_{m>m^\prime} &= \begin{pmatrix}
      0 & {g^{\alpha,>}_{mm^\prime}}^* & {g^{\alpha,<}_{mm^\prime}}^* & 0\\
      -g^{\alpha,>}_{mm^\prime} & 0 & 0 &  -g^{\alpha,<}_{mm^\prime}\\
     -g^{\alpha,>}_{mm^\prime}& 0 & 0 & -g^{\alpha,<}_{mm^\prime}\\
      0 & {g^{\alpha,>}_{mm^\prime}}^* & {g^{\alpha,<}_{mm^\prime}}^* & 0
      \end{pmatrix},
      \label{eq_nonintg1}
      \\
      \mathbf{G}^\alpha_{m=m^\prime} &= \frac{1}{2} \begin{pmatrix}
      0 & {g^{\alpha,>}_{mm}}^* & g^{\alpha,>}_{mm} & 0\\
      -g^{\alpha,>}_{mm} & 0 & 0 &  -{g^{\alpha,>}_{mm}}^*\\
     -g^{\alpha,>}_{mm}& 0 & 0 & -g^{\alpha,<}_{mm}\\
      0 & {g^{\alpha,>}_{mm}}^* & {g^{\alpha,<}_{mm}}^* & 0
      \end{pmatrix} ,
      \label{eq_nonintg2}
\end{align}
with 
\begin{align}
g^{\alpha,<}_{mm^\prime}(\omega) & \equiv -n^\alpha_\text{F}(\omega)\, e^{-i\omega(m-m^\prime)\delta t}\\
g^{\alpha,>}_{mm^\prime}(\omega) & \equiv \big(1-n^\alpha_\text{F}(\omega) \big)\, e^{-i\omega(m-m^\prime)\delta t}.
\end{align} 
These equations make the connection between $\mathbf{\Delta}_{mm^\prime}$ and
the standard textbook hybridization function ${\Delta}(\tau,\tau^\prime)$  [Eq.
(2) of the main text] manifest.  Equation~\eqref{eq_discreteDelta} allows to
use our formalism to compute impurity dynamics with an environment defined by
an arbitrary spectral density $\Gamma(\omega)$.

We emphasize that Eq.~\eqref{eq:IF_form} here represents the exact
discrete-time IF of a trotterized system, computed from its unitary Floquet
dynamics. Thus, in contrast to a brute-force discretization of the textbook
expression in Eq. (2) of the main text, Eq.~\eqref{eq:IF_form} produces a
physically meaningful evolution of the impurity [completely positive and trace
preserving (CPTP)] -- close by $\mathcal{O}(\delta t)$ to the exact
continuous-time Hamiltonian dynamics. Conversely, plugging a spectral density
$\Gamma(\omega)$ in Eq.~\eqref{eq_discreteDelta} and neglecting the
$\mathcal{O}(\delta t)$ corrections  generally
(slightly) breaks CPTP.

The impurity tensors $\hat D_m$ appearing in the overlap form are obtained directly from the Grassmann kernels corresponding to the impurity evolution: We convert each time-local impurity kernel at time step $m$ to an operator $\hat{D}_m$ that acts between a ``$\uparrow$'' two-fermion space (originally corresponding to the tensor product of input
and output Hilbert spaces of the $\uparrow$ impurity fermion) and a “$\downarrow$” two-fermion space (originally corresponding
to the tensor product of input and output Hilbert spaces of the $\downarrow$ impurity fermion), see e.g. superimposed red squares in Fig.~\ref{fig:Grassmann_labels}. Their tensor product, $\hat{D}_{\hat{O},t} = \hat{D}_1 \otimes \hdots \otimes \hat{D}_M$ (see main text), defines the product operator which we contract with the IF-MPS as shown in Fig.~\ref{fig:Grassmann_labels} and described in App.~\ref{app:mps}.

\section{Details on MPS computations}
\label{app:mps}
\renewcommand{\figurename}{Figure}
\renewcommand{\thefigure}{S\arabic{figure}}
  \begin{figure}[t]
\includegraphics[width=0.325\textwidth]{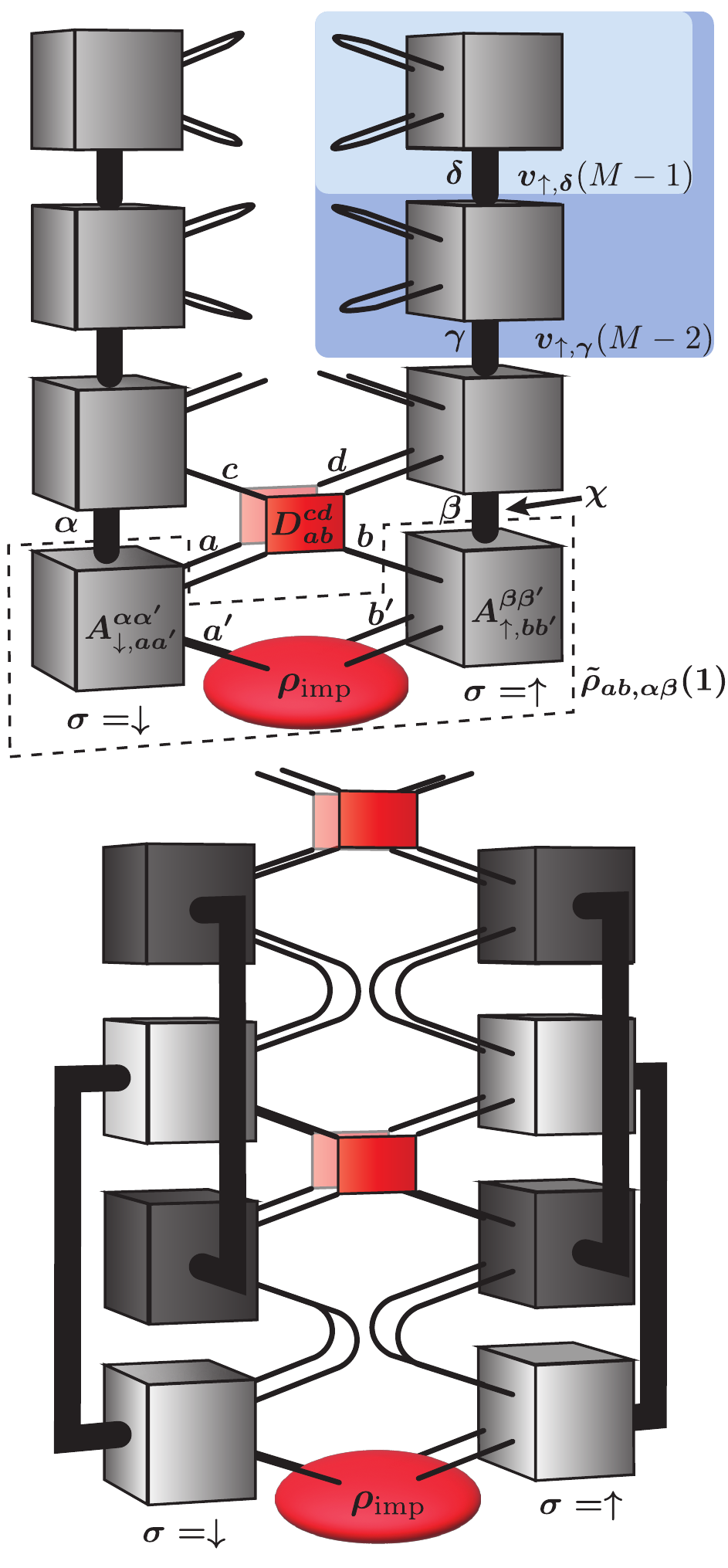}
\centering
    \caption{Top: Illustration of the IF-MPS contraction procedure for a single
    environment and two spin species. Pairs of impurity-legs on the forward and
    backward branch are combined and
    labelled by Latin letters, while virtual MPS indices are labelled by Greek letters. Each
    tensor $A$ of the IF-MPS can be viewed as a linear operator connecting two
    physical and two virtual spaces. The augmented density matrix
    $\tilde{\rho}_{ab,\alpha\beta}(1)$ for the first half time step is highlighted
    by dashed lines. The vectors $v_{\uparrow,\gamma}(M-1)$ and
    $v_{\uparrow,\beta}(M-2),$ necessary to extract the impurity density matrix
    from the augmented density matrices, are shaded blue. The whole tensor
    network corresponds to the vectorized impurity density matrix
    $\rho_{ef}(3)$ (after the action of IF$\rightarrow$impurity$\rightarrow$ IF ). Bottom: To compute non-equilibrium observables like the current---here for
    a two-terminal setup---we time evolve $\rho_\text{imp}(m)$ by contracting
    it successively with the IF-MPS of the right (light gray) and left (dark
    gray) environment before applying the impurity gate. In this example, we
    show the density matrix of the impurity after two full time steps.}
\label{fig:mps_ilustration}
\end{figure}

{\it Obtaining the MPS representation of the IF.}
 To obtain a MPS representation of the IF wave function $|I\rangle,$ introduced in Eq.~(
 5) of the main text, we apply a generalization of the Fishman-White (FW)
algorithm
that we adapted to BCS-like wave functions 
[52,57]. Its input is the two-point correlation matrix $\Lambda$ whose subblocks are given by \begin{equation}
    \Lambda_{\tau\tau^\prime} = \begin{pmatrix} \frac{\langle I | c_\tau c_{\tau^\prime}^{\dagger}|I\rangle}{\langle I|I\rangle} & \frac{\langle I | c_\tau c_{\tau^\prime}|I\rangle}{\langle I|I\rangle}\\
    \frac{\langle I | c^{\dagger}_\tau c^\dagger_{\tau^\prime}|I\rangle}{\langle I|I\rangle} &\frac{\langle I | c^{\dagger}_\tau c_{\tau^\prime}|I\rangle}{\langle I|I\rangle}\end{pmatrix},
    \end{equation}
where $\tau,\tau^\prime$ are points on the discretized Keldysh contour.
The FW algorithm encodes $\Lambda$ as a quantum circuit consisting of unitary gates. The accuracy of the circuit representation of $|I\rangle$ is set by an external parameter $\epsilon$ that is chosen beforehand. For $\epsilon\to 0,$ the circuit representation becomes exact. Applying the circuit to the vacuum (product) state of $4M$ spins yields the MPS representation of $|I\rangle.$

The ``local depth'' $D$  of the circuit scales logarithmically in evolution
time [42,49,52], with 
a prefactor that
increases as $\epsilon$ is decreased.
For numerical stability, it is advantageous to fix a very small
$\epsilon$ and continuously reduce the bond dimension $\chi$ of the MPS during the circuit contraction. For this, we use conventional singular value decomposition (SVD) where we fix
the maximal bond dimension $\chi_\text{max}$. 

%
In practice, we group all four tensor legs at equal time index $m$ into a single, larger physical Hilbert space of
dimension $d=2^4$; the bond dimension $\chi$ refers to these enlarged tensors
as depicted in Fig.~\ref{fig:mps_ilustration} (top).
This procedure yields a MPS that is proportional to the IF wave function, up to
errors through finite $\epsilon$ and SVD truncation errors.
To get the overall normalization $\langle I|I \rangle$, we are in principle free to insert arbitrary impurity evolution operators $U_\text{imp}$, imposing that the Keldysh partition function equals $1$. For environments defined by a continuous spectral density $\Gamma(\omega)$, however, different choices of $U_\text{imp}$ lead to slightly different normalizations, as time discretization slightly violates the CPTP property of the impurity evolution, as remarked in App.~\ref{app:overlap}. 
For the sake of efficiency, we choose to apply perfectly depolarizing channels on the impurity, which has the effect of connecting the forward and backward legs at a given variable index $\tau$\,
[65] (see upper half of top panel in Fig.~\ref{fig:mps_ilustration}). This allows us to minimize the cost of normalizing each IF separately.



Once the properly normalized IF-MPS have been obtained, we proceed as follows: The individual
tensors of the IF-MPS $\{A_\sigma(m)\}_{m=1}^M$ (gray bricks in Fig.~\ref{fig:mps_ilustration}) can formally be viewed as ``superoperators'' acting on the impurity with its physical legs and on the (compressed) environment with its virtual legs [66]. To obtain the full time evolution of the density matrix, we apply
these ``superoperators'' alternately with the local superoperator associated with the
 impurity Hamiltonian (red bricks in Fig.~\ref{fig:mps_ilustration}).


{\it Contracting with a single environment.}
First, let us consider the evolution of an impurity coupled to a single environment which is encoded
by two IF-MPS (one for each spin species). This setup is depicted in
Fig.~\ref{fig:mps_ilustration} (top) and corresponds to the calculations in Fig.~2 and
Fig.~4 of the main text. 

In the following notation, we use Latin letters to denote
combined indices of the forward and backward branch of the Keldysh contour. These indices correspond to
physical legs of the IF-MPS. Furthermore, we use Greek letters to label the virtual legs of the MPS. These represent a fictitious state of the
compressed environment associated with the MPS virtual bond space as depicted in Fig.~\ref{fig:mps_ilustration} (top). 

Our starting point is the vectorized initial density matrix $\rho_{ab}$ of the impurity,
where the indices $a,b$ correspond to the spin up- and down-
fermion respectively. We then add two additional indices, one for each IF, to
obtain an augmented density matrix $\tilde{\rho}(0)_{ab,\alpha\beta}$. Each time step consists of
i) the combined evolution of the impurity with the environment and ii) the evolution
of the impurity only. The former is formally expressed as:
\begin{equation}
  \tilde{\rho}_{ab,\alpha\beta}(2m+1) = A_{\downarrow,aa'}^{\alpha\alpha'}(m)A_{\uparrow,bb'}^{\beta\beta'}(m)\tilde{\rho}_{a'b',\alpha'\beta'}(2m),\label{eq:augmented_rho_if}
\end{equation}
where $A_{\sigma,aa'}^{\alpha\alpha'}(m)$ is the IF-MPS tensor at time step $m$.
For step ii), we apply the local impurity evolution represented by
the superoperator $D^{cd}_{ab}(m)$ (cf. Eq.~(
4) in main text):
\begin{equation}
  \tilde{\rho}_{ab,\alpha\beta}\big(2m+2\big) = D^{ab}_{a'b'}(m) \tilde{\rho}_{a'b',\alpha\beta}(2m+1). 
\end{equation}
To obtain the density matrix of the impurity from the augmented density
matrices at arbitrary intermediate times $\tilde{\rho}(m)$, we
recursively compute a set of vectors $v_{\sigma,\alpha}(m)$ for each of the two IF (corresponding to $\sigma = \uparrow$ and $\sigma =\downarrow,$ respectively):
\begin{equation}
  v_{\sigma,\alpha'}(m-1) = A^{\alpha\alpha'}_{\sigma,aa'}(m) v_{\sigma,\alpha}(m) \frac{1}{2}\mathds{1}_a\mathds{1}_{a'},\label{eq:trace_environment}
\end{equation}
with $v_{\sigma,\alpha}(M)=1$ and $\mathds{1}_a$ being vectorized identities, see upper panel of Fig.~\ref{fig:mps_ilustration}.
These vectors allow to ``trace out the environment'' at intermediate times: The density matrix of the impurity $\rho(m)$ can then be obtained from the
augmented density matrices as
\begin{align}
  \rho_{ab}(2m) &= v_{\downarrow,\alpha}(m)v_{\uparrow,\beta}(m) \tilde{\rho}_{ab,\alpha\beta}(2m),\\
  \rho_{ab}(2m-1) &= v_{\downarrow,\alpha}(m)v_{\uparrow,\beta}(m) \tilde{\rho}_{ab,\alpha\beta}(2m-1).
\end{align}
Physically, this is equivalent to completing the Keldysh contour to the final
time $T$ by evolving the impurity with the perfectly depolarizing channel for
each timestep later then $m$. Numerically, this is slightly more convenient than using identities as in Eq.~\eqref{eq:U_imp}, as the matrices to be multiplied in Eq.~\eqref{eq:trace_environment} are smaller.

{\it Computation of the current.}
To compute the current as in Fig.~3 of the main text, it is necessary to
compute the  IF of the left and right environment separately, such that we
can access the impurity density matrix before and after the individual interaction with the left
and with the right environments, as depicted in the bottom panel of Fig.~\ref{fig:mps_ilustration}. This leads to a modified prescription for time evolution including four independent IF: The augmented density matrix has now four
environment indices corresponding to the virtual bond of each of the four IF.
For each timestep, we perform the operation in Eq.~\eqref{eq:augmented_rho_if}
twice, once for the left and once for the right environment IF. To extract the
impurity density matrix from the augmented density matrix at each substep, all
of the additional indices have to be contracted with the corresponding vector
$v_{L/R,\uparrow/\downarrow,\alpha}.$ This vector is obtained individually for each
IF according to Eq.~\eqref{eq:trace_environment}. The required computer memory
of the contraction algorithm scales thus as $O(16 \chi^4),$ where $\chi$ is the
maximal bond dimension of each IF. While this requirement of memory resources imposes a bound on manageable bond dimensions in practice, we demonstrate in Fig.~3 that even bond dimensions as low as $\chi=32$ yield converged results that are competitive with state-of-the-art methods for non-equilibrium dynamics.